\documentclass[aps,superscriptaddress,twocolumn,superscriptaddress,showpacs,showkeys,
groupedaddress,nofootinbib,nobalancelastpage,nobibnotes]{revtex4}
\pdfoutput=1
\usepackage{graphicx,color}
\usepackage{amssymb}
\usepackage{epstopdf}
\usepackage{amsmath}
\usepackage{multirow}
\usepackage{comment,color}
\newcommand{\refmod}[1]{{\leavevmode\color{black}#1}}
\newcommand{\phigal}{\phi_{\text{gal}}}
\newcommand{\phiatmo}{\phi_{\text{atmo}}}
\newcommand{\phisfg}{\phi_{\text{X-pp}}}
\newcommand{\phitde}{\phi_{\text{X}-\text{p}\gamma}}
\newcommand{\fxpp}{F_{\text{X-pp}}}
\newcommand{\fxpg}{F_{\text{X}-\text{p}\gamma}}

\newcommand{\ie}{{\it i.e.}}

\newcommand{\eg}{{\it e.g.}}

\newcommand{\cf}{{\it cf.}}

\newcommand{\eq}{Eq.}

\newcommand{\fig}{Fig.}
\newcommand{\Fig}{Fig.}

\newcommand{\Ref}{Ref.}
\newcommand{\Refs}{Refs.}
\newcommand{\Sec}{Section}

\newcommand{\App}{Appendix}

\newcommand{\Tab}{Tab.}
\newcommand{\Tabs}{Tabs.}

\newcommand{\equ}[1]{\eq~(\ref{equ:#1})}
\newcommand{\figu}[1]{\fig~\ref{fig:#1}}

\newcommand{\bi}{\begin{itemize}}
\newcommand{\ei}{\end{itemize}}

\usepackage{hyperref}

\begin{document}

\title{A Multi-Component Model for the Observed Astrophysical Neutrinos}

\author{Andrea Palladino}
\email{andrea.palladino@desy.de}
\affiliation{Deutsches Elektronen-Synchrotron (DESY), Platanenallee 6, D-15738 Zeuthen, Germany}
\author{Walter Winter}
\email{walter.winter@desy.de}
\affiliation{Deutsches Elektronen-Synchrotron (DESY), Platanenallee 6, D-15738 Zeuthen, Germany}

\date{}
\begin{abstract}
 We propose a multi-component model for the observed diffuse neutrino flux, including the residual atmospheric backgrounds, a Galactic contribution (such as from cosmic ray interactions with gas), an extra-galactic contribution from $pp$ interactions (such as from starburst galaxies) and a hard extragalactic contribution from photo-hadronic interactions at the highest energies (such as from Tidal Disruption Events or Active Galactic Nuclei). We demonstrate that this model can address the key problems of astrophysical neutrino data, such as the different observed spectral indices in the high-energy starting and through-going muon samples, a possible anisotropy due to Galactic events, the non-observation of point sources, and the constraint from the extragalactic diffuse gamma-ray background. Furthermore, the recently observed muon track with a reconstructed muon energy of 4.5~PeV might be interpreted as evidence for the extragalactic photo-hadronic contribution. We perform the analysis based on the observed events instead of the unfolded fluxes by computing the probability distributions for the event type and reconstructed neutrino energy. As a consequence, we give the probability to belong to each of these astrophysical components on an event-to-event basis.
\end{abstract}

\maketitle

\maketitle


\section{Introduction}

In the last few years, the IceCube experiment has opened a new era in neutrino astronomy, providing the first evidence for high-energy astrophysical neutrinos~\cite{Aartsen:2013jdh}. The analysis used by IceCube to claim the discovery relies on events with the interaction vertex contained in the inner detection volume, also known as high energy starting events (HESE)~\cite{Aartsen:2013jdh,Aartsen:2014gkd,Aartsen:2015knd}. For this analysis, a veto based on the outer detector layer has been implemented to reduce the impact of atmospheric down-going neutrinos and muons; for instance, it rejects (mainly muon) neutrinos accompanied by an atmospheric muon.
The HESE analysis includes both muon tracks and showers (cascades) from both the Southern and Northern hemispheres, with slightly different sensitive energy ranges.
An alternative used before the HESE discovery -- and re-vived recently -- is the detection of through-going muons originating from interactions of $\nu_\mu$ from the Northern hemisphere (which are free of atmospheric muon backgrounds) outside the detector~\cite{Aartsen:2015rwa,Aartsen:2016xlq}. 
Similar analysis techniques are  used in ANTARES \cite{Collaboration:2011nsa} and the coming KM3NeT experiment  \cite{Adrian-Martinez:2016fdl}. 

The origin of the observed astrophysical neutrinos remains a mystery. While the directional event distribution is isotropic at a first glance -- pointing towards an extragalactic origin of most of the astrophysical neutrinos -- a more detailed analysis reveals a $\sim 2 \sigma$ excess close to the Galactic plane when only HESE events with deposited energy above 100 TeV are considered~\cite{Aartsen:2017mau}; see \eg\ \Refs~\cite{Razzaque:2013uoa,Ahlers:2013xia,Neronov:2015osa, Palladino:2016zoe, Palladino:2016xsy} for recent discussions. 
Concerning the spectrum, the through-going muon dataset suggests an hard spectrum close to $E_{\nu}^{-2}$ where the HESE sample is described by a much softer spectrum, close to $E_{\nu}^{-3}$~\cite{Aartsen:2017mau}. Consequently, the single power law hypothesis of the astrophysical flux has been questioned~\cite{Wang:2014jca,Palladino:2016xsy,Chen:2014gxa,Marinelli:2016mjo,Chianese:2017jfa,Borah:2017xgm}.

Stacking searches using gamma-ray catalogues of popular candidate classes, such as Gamma-Ray Bursts (GRBs)~\cite{Aartsen:2017wea} and Active Galactic Nuclei (AGN) blazars~\cite{Aartsen:2016lir}, have revealed that the observed flux cannot be dominated by these object classes. For example, 
the contribution from blazars has been found to be  smaller than about $\sim 20\%$ of the observed diffuse flux~\cite{Aartsen:2016lir}; see also \cite{Palladino:2017aew,Wang:2015woa}. 

More generically, no point sources have been resolved so far, which indicates that most of the observed extragalactic neutrinos ought to come from abundant sources with low luminosities~\cite{Kowalski:2014zda,Ahlers:2014ioa,Murase:2016gly}. If a large portion of the energy range is to be described, these neutrinos have to follow a power law with a spectral index between about $E_{\nu}^{-2}$ and  $E_{\nu}^{-2.5}$ as expected from Fermi shock acceleration for the primary cosmic rays. Such a neutrino spectrum following the primary spectrum is expected for $Ap$ or $pp$ interactions. 

Starburst galaxies are an ideal source candidate for that~\cite{Loeb:2006tw,Tamborra:2014xia,Chang:2014sua} (we will discuss other alternatives later), which however face two other challenges: a) the neutrino spectrum must not be much softer than $E^{-2}$ to avoid the constraint from the observed diffuse extragalactic gamma-ray background~\cite{Murase:2013rfa,Bechtol:2015uqb} and b) cosmic-rays accelerated in starburst galaxies are unlikely to produce neutrinos with energies much larger than PeV energies. However a muon track with a reconstructed energy of 4.5 PeV and a likely neutrino energy $\gtrsim$ 10~PeV has been recently detected by IceCube \cite{Aartsen:2016xlq} -- which requires a 100~PeV primary. 

Such high-energy events could come from rarer sources producing neutrinos by $A\gamma$ or $p\gamma$ interactions, which typically have hard spectra which can easily peak at PeV energies and beyond, see \eg\ \Ref~\cite{Winter:2013cla}. One example, which has been recently drawing a lot of attention, are neutrinos from (jetted) Tidal Disruption Events (TDEs)~\cite{Wang:2011ip,Wang:2015mmh,Dai:2016gtz,Senno:2016bso,Lunardini:2016xwi,Biehl:2017hnb,Guepin:2017abw},
which we will use as a prototype. Compared to other sources with similar spectral properties, such as AGN blazars, we will see that the TDE hypothesis can be tested by multiplet searches in the near future. In addition, a self-consistent description with ultra-high energy cosmic rays (UHECRs) can be performed~\cite{Biehl:2017hnb}, \ie, these neutrinos may come from the sources of the cosmic rays at the highest energies. Note that in this case the primaries include nuclei  heavier than protons, as indicated by recent results of the Pierre Auger collaboration~\cite{Aab:2016zth}.

Since it is clear that a single (astrophysical) flux component cannot address all open questions, we propose a multi-component model to draw a self-consistent picture of the observed astrophysical neutrino flux which can satisfy all of these constraints. Of course, a relevant question is how many components do we really need -- which we are going to address quantitatively.
Compared to many earlier works, we start off with the information on the observed events and reconstruct the probability distribution for the reconstructed neutrino energy and interaction type; see \Refs~\cite{Vincent:2016nut,DAmico:2017dwq} for other such examples.  

The structure of the paper is as follows: In \Sec~\ref{sec:methods}, we present the most recent IceCube dataset, namely the six years HESE dataset and the eight years through-going muon dataset, and we  describe our procedure to obtain the probability distribution function of the reconstructed neutrino energy; for technical details, see \App~\ref{app:methods}. We furthermore present in \Sec~\ref{sec:multi-component} our multi-component model, analyzing separately the contributions of atmospheric neutrinos, Galactic neutrinos, extragalactic $pp$ neutrinos (such as neutrinos from starburst galaxies) and extragalactic $p\gamma$ neutrinos (such as neutrinos from TDEs or AGNs) from a theoretical point of view. We show the results of the fit in \Sec~\ref{sec:results}, where we also discuss how many components are really needed, and what we can learn in the future. An event-based assignment to the different components (probabilities)  can be found in \App~\ref{app:prob}. Finally, we summarize in \Sec~\ref{sec:conc}.

\section{Methods}
\label{sec:methods}

In this section, we describe the used datasets and our energy reconstruction technique. 

\subsection{The IceCube datasets}
\label{sec:dataset}

Here we introduce the latest two datasets provided by the IceCube collaboration after six years and eight years of data taking~\cite{Aartsen:2017mau}, which we use for our analysis: high energy starting events (HESE) dataset and the through-going muon dataset. 

\subsubsection*{High Energy Starting Events (HESE)}
\label{sec:hese}

\begin{figure}[t]
\begin{center}
\includegraphics[width=\columnwidth]{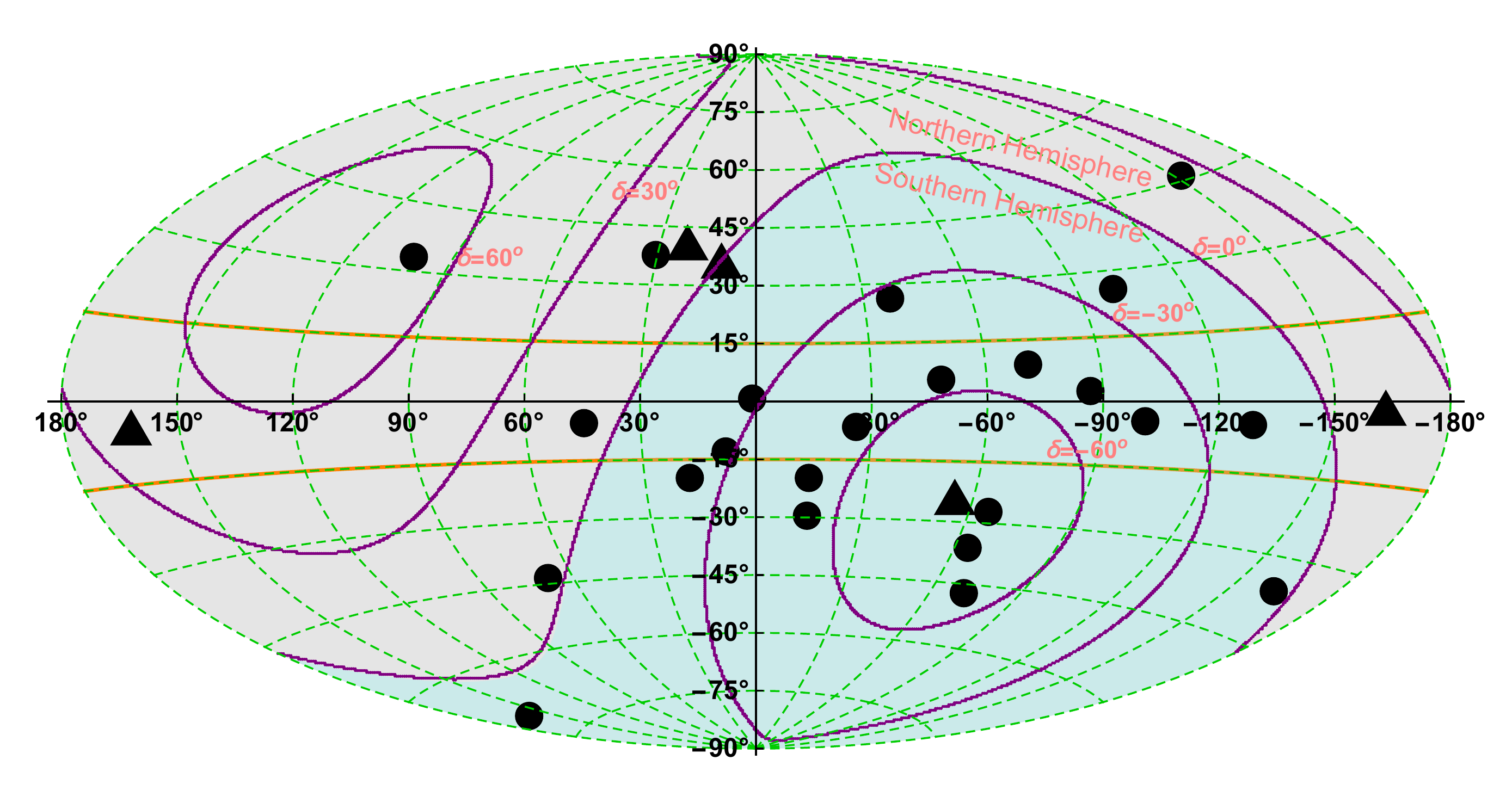}
\end{center}
\caption{HESE events with deposited energy above 100 TeV in in Galactic coordinates (circles: showers, triangles: tracks).  The purple contours represent intervals of declinations separated by $30^\circ$, whereas the brown lines denote the Galactic latitudes $b=15^\circ$ and $b=-15^\circ$. Taking into account that the average angular resolution of shower-like events is $\sim 15^\circ$ we notice an accumulation of events close to the Galactic plane, \ie, within the brown curves.}
\label{fig:map}
\end{figure}

The HESE are the  events with neutrino interaction vertex contained in the detector volume; this analysis has been used by IceCube to claim the evidence of an extraterrestrial flux of high energy neutrinos \cite{Aartsen:2013jdh}. 
The most recent dataset includes 2078 days (5.7 years) of operation with 82 HESE detected~\cite{Aartsen:2017mau}: 22 muon tracks, 58 showers and two events produced by a coincident pair of background muons from unrelated cosmic ray air showers, that have been excluded from the analysis.
These events are characterized by a deposited energy larger than 30 TeV, and the most energetic HESE has deposited an energy of 2~PeV into the detector. 

The flux attributed to astrophysical neutrinos has been, to a first approximation, described by a power law spectrum and an isotropic distribution. The per-flavor flux has been given by
\begin{equation}
\frac{d\phi_\ell^{\text{\text HESE}}}{dE}= F_\ell \times \frac{10^{-18}}{\rm GeV \ cm^{2} \ s \ sr } \left(\frac{E}{\rm 100 \ TeV} \right)^{-\alpha}
\label{eq:hese}
\end{equation}
with $F_\ell=2.5 \pm 0.8$ and $\alpha=2.92^{+0.33}_{-0.29}$ \cite{Aartsen:2017mau}.\footnote{In IceCube 
analyses, usually  the equipartition of the flavors is assumed, which is roughly expected from high energy neutrinos produced by pion decays after flavor mixing.}
The HESE dataset is dominated by events coming from the Southern hemisphere because above several hundred TeV the absorption probability of neutrinos crossing Earth becomes greater than 50\%~\cite{Aartsen:2017kpd}. In addition, re-call that the veto method of atmospheric neutrinos only works for down-going neutrinos, \ie, for neutrinos from the Southern hemisphere. 

The HESE are isotropically distributed when the full dataset is considered. On the other hand, when the analysis is limited to events above $\sim$100~TeV, an anisotropy appears, with an accumulation of events close to the Galactic plane which is visible by bare eye; see \figu{map}. The present statistical significance of this accumulation of events is about 2$\sigma$.\footnote{For one (isotropic) event, the probability to end up in the Galactic region is equal to $\sin b \simeq 0.26$. 
Using this information, we derive the probability that a certain number of events is contained in this region, similarly to the calculation performed in \Ref~\cite{Palladino:2017aew}. We obtain that the probability that four or more events are found in the Galactic region is 72\%, whereas the probability that eight events or more are found there is only 7\% (which translates into a 93\% confidence level, or, indeed $1.8\sigma$ for the eight events found there). }
This slight excess of events close to the Galactic plane was already present in older IceCube datasets and has been already discussed in \Refs~\cite{Neronov:2015osa,Palladino:2016zoe,Palladino:2016xsy,Pagliaroli:2017fse}. In this work, we will also take into account the possible existence of a Galactic component of high energy neutrinos.

\subsubsection*{Through-going muons (TGM)}
\label{sec:muon}

The IceCube collaboration has collected a sample of charged current events produced by up-going muon neutrinos \cite{Aartsen:2017mau} from 2009 to 2017 .
The through-going muons come from muon neutrinos interacting outside the detector. Thus, in order to avoid the atmospheric muon background, the field of view must be restricted to the Northern hemisphere such that atmospheric muons are absorbed in Earth.

The high energy sample (with muon reconstructed energy above $\sim$ 200 TeV) contains 36 such events; 
a purely atmospheric origin of them is excluded at more than 5$\sigma$. 
The most energetic event corresponds to a reconstructed muon energy of 4.5~PeV.

The corresponding cosmic muon neutrino and antineutrino flux has been obtained with a power law fit to the data:
\begin{equation}
\frac{d\phi_\mu^{\text{\text TGM}}}{dE}=F_\mu \times \frac{10^{-18} }{\rm GeV \ cm^{2} \ s \ sr } \left(\frac{E}{\rm 100 \ TeV} \right)^{-\alpha} \, .
\label{eq:muon}
\end{equation}
The parameters are $F_\mu=1.01^{+0.26}_{-0.23}$ and $\alpha=2.19 \pm 0.10$. This analysis is sensitive mostly to muon neutrinos and antineutrinos plus a small contribution from the $\tau$-leptons that decay into muons. 

The spectral discrepancy between the HESE and TGM datasets has a significance of about 3$\sigma$ and will be  investigated in this work.

\subsection{Neutrino energy reconstruction}
\label{sec:montecarlo}

Here we summarize our energy reconstruction mechanism, for details, see \App~\ref{app:methods}.

As the general ansatz we start with the list of events, sorted by topology (shower-like or track-like) and dataset (HESE or TGM); see \Tabs~\ref{tab:shower}, \ref{tab:track}, and \ref{tab:muon}. For each of these events, apart from the topology, the sky position (and its uncertainty) is known, as well as the energy deposited in the detector in form of secondary particles and, eventually, light. 
As initial information on the energy of  each event, from here on we always use: \\
- the deposited energy for shower like events; \\
- the reconstructed muon energy for TGM events.\\
 The list of events contains a residual background from atmospheric neutrinos and muons passing the veto.

The final task will be to determine for each event $i$ the reconstructed neutrino energy distribution $C_i(E_\nu)$ including  the probability that a shower was created from a $\nu_e$, $\nu_\tau$ or neutral current interaction, or a track was created from a $\nu_\mu$ or $\nu_\tau$ induced muon. Our Monte Carlo procedure to obtain this probability distribution is described in detail in \App~\ref{app:methods}. Note that (in terminology) we go from incident (true) neutrino energy $E_\nu$ to the deposited energy $E_{\text{dep}}$ in the detector, which leads to a distribution for the reconstructed neutrino energy $E_\nu$ -- which we label the same as the incident energy for the sake of readability.

This reconstructed energy distribution can be used to construct binned error bars on the flux in terms of the reconstructed neutrino energy for arbitrary datasets if one deconvolves the total event rate in each bin with the effective area. This leads to similar results compared to the IceCube procedure at the highest energies (see \eg\ data points in \figu{comp}); however,  the residual atmospheric background will show up explicitly at low energies. As another difference, we separate tracks and showers since the track sample is more affected by atmospheric backgrounds than the shower sample, which will be relevant for the construction of our multi-component model.

\section{The multi-component model}
\label{sec:multi-component}
\begin{table*}[t]
\begin{center}
\begin{tabular}{llrrlrp{4cm}}
\hline
Name & Production mechanism &  Normalization & $\alpha$ & Angular & Energy range    & Examples/Comments \\
\hline
{\bf Atmospheric} & Pion/kaon decay, atm. muons & $F^\mu_{\text{atmo}}$, $F^e_{\text{atmo}}$ & 3.7 & Isotropic  & $\lesssim$ 0.2-0.5 PeV  &  \\
 & Charmed meson decays (prompt) & (fixed) & 2.7 & Isotropic  & (never) & Norm. from \cite{Aartsen:2017mau} (upper limit) \\
{\bf Galactic} & $Ap$ or $pp$ interactions & $F_{\text{gal}}$ & $\sim 2.6$ &  Galactic & $\lesssim$ PeV & {\bf Cosmic ray interactions with gas}, point sources \\
{\bf X$_{\text{pp}}$} (extragal.) & $Ap$ or $pp$ interactions &  $F_{\text{X-pp}}$ & $\sim 2$ & Isotropic  & 0.2 - 2 PeV &  {\bf Starburst galaxies}, radio galaxies, AGN winds, halo/galaxy mergers \\
{\bf X$_{\text{p}\boldsymbol{\gamma}}$} (extragal.)  & $A\gamma$ or $p\gamma$ interactions & $F_{\text{X-p}\gamma}$  & $\ll  2$ &  Isotropic & $> 2$ PeV  & {\bf Tidal Disruption Events (TDEs)}, AGN blazars, LL-GRBs \\
\hline
\end{tabular}
\end{center}
\caption{Summary of the main characteristics of the different components that are present in our multi-component model.  The column ``Normalization'' refers to the (free) normalization parameter used in the fit (unless it is fixed), the column ``$\alpha$'' to the approximate spectral index of the incident neutrino flux $\propto E_\nu^{-\alpha}$, the column ``Energy range'' to the approximate energy range where the spectrum is found to dominate (which depends  on the sky direction and event topology), and the column ``Angular'' to the assumed rough angular distribution. In the last column, our baseline model for the spectral shape is marked boldface.
\label{tab:summarize}}
\end{table*}%

\normalsize

In this section we introduce our model, which includes \textit{(i)} the atmospheric background (conventional and prompt neutrinos, atmospheric muons), since we will include it in the reconstructed data points; \textit{ii)} Galactic neutrinos generated in $pp$ collision in the Galactic disk plus Galactic neutrinos from point sources; \textit{iii)} extragalactic neutrinos from $pp$ (or $Ap$) interactions, such as neutrinos from starburst galaxies; and \textit{iv)} neutrinos from $p\gamma$ (or $A\gamma$) interactions, such as neutrinos produced in TDEs, AGNs, or GRBs. We summarize these different components in \Tab~\ref{tab:summarize}.

\subsection{Residual atmospheric backgrounds}

Atmospheric neutrinos and atmospheric muons passing the veto system produce a residual background for the  high energy neutrino detection. The conventional atmospheric (neutrino and muon) background follows the atmospheric muon spectrum $\propto E^{-3.7}$ (determined by the initial cosmic-ray spectrum modified by interactions of pions and kaons), whereas prompt neutrinos at higher energies are characterized by an $E^{-2.7}$ spectrum (following the initial cosmic-ray spectrum because the secondaries decay faster than they interact). 
The flavor composition of atmospheric neutrinos is to a  good approximation $(\nu_e:\nu_\mu:\nu_\tau) \simeq (0:1:0)$ in the first case, where pion decays dominate and kaon decays give a smaller but not negligible contribution. In pion decays, muon neutrinos and muons are produced. The secondary muons have no time to decay before reaching the surface of the earth, therefore the conventional background is essentially given by $\nu_\mu$ and $\bar{\nu}_\mu$.
As far as the prompt neutrinos are concerned, the flavor composition is given by $(\nu_e:\nu_\mu:\nu_\tau) \simeq (1:1:0)$ since charmed meson decays dominate here. 

Atmospheric muons contribute to the background from the Southern hemisphere (down-going events). Note that we will not consider the actual atmospheric fluxes, but a residual background passing the veto systems with a normalization describing observations.

The HESE track sample is much more affected by the atmospheric backgrounds than the HESE shower sample because atmospheric muons dominate this background (in the Southern hemisphere) and the atmospheric electron neutrino flux, leading to showers, is suppressed compared to the muon neutrino flux (in the Northern hemisphere). In order to avoid a bias on the shower sample, we assume different backgrounds for the $\phi_\mu$ and $\phi_e$ fluxes in our model, reflecting the different systematics of these topologies:
\begin{equation}
\label{eq:atmo}
\begin{split}
\frac{d \phiatmo^{e,\mu}}{dE_\nu}  =\frac{10^{-18}}{\rm GeV \ cm^2 \ s \ sr} \bigg[ F_{\text{atm}}^{e,\mu} \left(\frac{E_\nu}{100 \, \text{TeV}} \right)^{-3.7} +  \\ 
F_{\text{prompt}}\left(\frac{E_\nu}{100 \, \text{TeV}} \right)^{-2.7} \bigg] \, .
\end{split} 
\end{equation}
Here $F_{\text{atm}}^e$ and $F_{\text{atm}}^\mu$ are two free (systematics) parameters of the model, whereas we fix $F_{\text{prompt}}=0.66$ to the present upper limit of the prompt neutrino flux per flavor \cite{Aartsen:2017mau}.
This choice is motivated as conservative background estimate, consistent with the total background expected for shower-like events in IceCube~\cite{Aartsen:2017mau}.
We will see later that the prompt contribution does not affect our results because it is about an order of magnitude smaller than the astrophysical contributions.

  Let us re-call that the background expected by IceCube is  in tension with the observations, as they expect $33.9^{+9.3}_{-6.4}$ background tracks (see Tab.~3 of \cite{Palladino:2017qda}), whereas only 22 tracks have been observed~\cite{Aartsen:2017mau}. Note that in addition the astrophysical signal contributes to the track-like events (about 20\% of the astrophysical neutrinos should produce tracks, assuming pion decay as production mechanism \cite{Palladino:2015zua}). This implies about 40 tracks are expected instead of the 22 observed, which represents a discrepancy  of about $(40-22)/\sqrt{22} \sigma \simeq 3.8 \sigma$; this tension between expectation and observation was already present in the older datasets (see \eg\ Tab.~4 of \cite{Aartsen:2014gkd} for the three-year dataset). 

In our analysis, the two systematics parameters $F^\mu_{\text{atmo}}$ and $F^e_{\text{atmo}}$ are to be marginalized over, which means that no particular input is assumed for these. We will verify that, at the end,  we will reproduce the total background event expected by IceCube~\cite{Aartsen:2017mau}, namely $40.8^{+13.5}_{-8.3}$ events, reasonably well. While this procedure does not affect our result at the highest energies or for the astrophysical components qualitatively, the obtained values for the systematics parameters can be interpreted  with respect to a possible mis-identification of events and with respect to the possible origin of the discussed track discrepancy. 

\subsection{Galactic component}
\label{subsec:gal}

The presence of a Galactic component, that could affect especially the events coming from the Southern hemisphere, has been already proposed in \cite{Neronov:2015osa,Palladino:2016zoe,Palladino:2016xsy,Denton:2017csz,Marinelli:2016mjo}, based on the IceCube observations. A guaranteed (diffuse) flux comes from the interactions of cosmic rays with gas; however, this contribution is typically small in models in which the local cosmic ray density is assumed to be representative for the whole Milky Way~\cite{Ahlers:2013xia,Joshi:2013aua}. If, however, a possible inhomogeneous cosmic ray distribution in our galaxy is taken into account, higher contributions can be obtained. In \cite{Pagliaroli:2016lgg} this flux has been calculated taking into account different cosmic ray distributions;
 the most optimistic case corresponds to the so-called KRA$\gamma$ model~\cite{Gaggero:2015xza}, in which the spectral index of cosmic rays is a function of the distance from Galactic center, namely harder close to the Galactic center and softer far away from it. In this scenario the expected number of HESE in IceCube are about six  after 5.7 years of exposure. The additional contribution from Galactic point sources
is considered in \cite{Pagliaroli:2017fse}. The expected number of events from point sources is $\sim 3$ with the present exposure. Therefore the total expected number of Galactic events must not exceed 9 HESE in about 6 years of detection. 
These theoretical expectations are perfectly compatible with the most recent experimental limits on the Galactic flux provided by ANTARES \cite{Albert:2017oba} and the IceCube \cite{Aartsen:2017ujz}. 

Following these considerations, we propose a Galactic flux that is in agreement with the present knowledge. Considering that the observed flux of Galactic cosmic rays observed at Earth is $E^{-2.7}$, we would expect that high energy neutrinos produced by $pp$ interactions between Galactic cosmic rays and gas  will follow the power law spectrum of the primary particles -- being somewhat harder from the interactions $\propto E^{-2.6}$~\cite{Kelner:2006tc}. We assume this shape in our multi-component model. Note that  in a more aggressive scenario, such as the KRA$\gamma$ model, the neutrino spectrum could be harder. \refmod{An harder galactic neutrino component has been analyzed in \cite{Palladino:2016xsy}. This choice would reduce the discrepancy between what is observed from South and from North, increasing of a few \%  the normalization of our \lq\lq residual atmospheric background\rq\rq. Since we are looking for a Galactic component that can emphasize the North-South spectral asymmetry, we assume a softer spectrum in our model. Let us remark that we will check the compatibility with the present constraints (from both theory and experiment) at the end of the calculations.} 

We use an energy cutoff in the neutrino spectrum at the corresponding knee of the cosmic rays at about 3 PeV for protons and, correspondingly, 150 TeV for neutrinos\footnote{We assume an exponential energy cutoff for the proton spectrum, which implies, approximately, the square root of the exponential cutoff for the neutrino spectrum~\cite{Kelner:2006tc}.}, considering that the average energy of neutrinos from $pp$ interaction is 5\% of the primary proton energy \cite{Kelner:2006tc}. This cutoff is compatible with recent multi-component models for cosmic rays at around the knee~\cite{Gaisser:2013bla} (see Fig.~1 in \Ref~\cite{Joshi:2013aua} for the neutrino flux), assuming that the proton component dominates the neutrino production; it is also compatible with theoretical arguments on the maximal acceleration energy of Galactic cosmic rays~\cite{Ellison:1997an}.

Consequently, the diffuse Galactic flux of high energy neutrinos is assumed to follow
\begin{equation}
\begin{split}
\frac{d\phigal}{dE_\nu}= \frac{F_{\text{gal}} \times 10^{-18}  }{\rm GeV \ cm^2 \ s \ sr}  \left(\frac{E_\nu}{100 \, \text{TeV}} \right)^{-2.6} \times \\
\exp 
\left(-\sqrt{\frac{E_\nu}{150 \, \text{TeV}}} \right) \,
\end{split}
\end{equation}
in which the only free parameter is the normalization $F_{\text{gal}}$.
In order to satisfy the present theoretical limits, we evaluate the expected number of events as
\begin{equation}
N_{\text{gal}}^\ell= 4 \pi  \ T  \ \int_0^\infty \frac{d\phigal}{dE_\nu} \times A_{\text{eff}}^\ell \ \times dE_\nu \, ,
\label{equ:nev}
\end{equation}
using the average HESE effective area~\cite{Aartsen:2013jdh}. This is a good choice since different intervals of terrestrial declination are covered in the inner Galaxy region, namely from $-90^\circ \leq \delta \leq 30^\circ$, see also \Fig~\ref{fig:map}.\footnote{We have checked that using the Southern sky effective area instead of the all-sky one, the result is only marginally affected; the expected number of events is 10\% larger considering an $E^{-2.6}$ spectrum with the energy cutoff at 150 TeV. This is naturally expected by the fact that -- for a soft spectrum like this -- neutrinos around 100 TeV provide the largest contribution to the events. At this energy, the effective areas are almost independent of declination -- whereas they are very different at higher energies, where the absorption of neutrinos in Earth becomes relevant, penalizing neutrinos coming from Northern hemisphere.} 
We obtain  from \equ{nev} that $N_{\text{gal}} \simeq 8.4 \times F_{\text{gal}} \times T/(5.7 \, \text{yr})$, which implies that $F_{\text{gal}} \lesssim 1.1$ in order not to overshoot the 9 Galactic events expected in 6 years considering the most optimistic model, as described above.
We will verify if our final result is in agreement with this constraint in the following of the paper.

Finally, note that we have multiplied \equ{nev} with the full solid angle $4\pi$, which does not mean that the Galactic flux is present in the whole sky. This choice however allows for an easy comparison with the  IceCube measurement, for which usually the average flux is given per steradian. 
In our model we take into account that the Galactic flux is present in the region of Galactic latitude $|\varphi| \leq 5^\circ$ and of longitude $| \lambda | \leq 45^\circ$ (inner galaxy), whereas it is not present outside. 
Therefore we can write the normalization in an alternative manner for the Galactic components, as follows:
$$
\frac{d \tilde \phi_{\text{gal}}}{dE_\nu} = 
\left\{
\begin{array}{lllll}
 & \frac{d\phigal}{dE_\nu} \frac{4\pi}{\Omega}  & & \mbox{for }  |\varphi| \leq 5^\circ \mbox{ and } | \lambda | \leq 45^\circ \\
 & 0 & & \mbox{for }  |\varphi| > 5^\circ \mbox{  or  } | \lambda | > 45^\circ \\
\end{array}
\right. \, ,
$$
where $\Omega = 4 \sin |\varphi_{\text{max}}| \cdot |\lambda_{\text{max}}|$, with $|\varphi_{\text{max}}| = 5^\circ$ and $|\lambda_{\text{max}}| = 45^\circ$. 

\subsection{Extragalactic $\boldsymbol{pp}$ neutrinos ($\boldsymbol{\text{X}_{\text{pp}}}$)}

Neutrinos are created via $pp$ (or $Ap$) interactions in astrophysical environments rich of gas, which serves as target for the interactions. In this case the accelerated primaries with a non-thermal spectrum collide with thermal protons, producing about equal amounts of $\pi^+$, $\pi^0$ and $\pi^-$ leading to a neutrino spectrum adopting the spectral shape of the non-thermal primaries -- which is typically a power law with a spectral index of about two.

Our prototype candidate class are Starburst Galaxies (SBGs), although this is not the only possible source class.
A SBG is a galaxy that has  an extremely high star formation rate of the order 
of $10-100 \  M_\odot$/year, while normal galaxies have $1-5 \ M_\odot$/year \cite{Kewley:2001ng}. 
A SBG is full of young stars, which implies that supernova events are frequent with a rate of the order of $0.03 -0.3$ per year; moreover, it must contain a large amount of gas available to be injected into the star-forming processes. The simultaneous presence of cosmic rays, accelerated in shock waves created by the exploding core-collapse SNe, and targets, \ie, large amounts of gas,  makes SBGs promising sources of cosmic neutrinos produced by $pp$ interactions. SBGs are mainly observed in the infrared band because the bulk of their emission (UV radiation emitted by young stars) is absorbed and re-emitted in this band. 

The SBGs have been proposed in several papers as source of IceCube neutrinos \cite{Loeb:2006tw,Tamborra:2014xia,Chang:2014sua} although they cannot explain the entire IceCube signal \cite{Bechtol:2015uqb,Murase:2016gly}. There have been many other alternatives discussed in the literature, which potentially also fit this category, such as radio galaxies~\cite{Tavecchio:2017utw}, AGN winds~\cite{Liu:2017bjr}, and halo/galaxy mergers~\cite{Yuan:2017dle}.

In our model, we assume an $E^{-2}$ spectrum with an energy cutoff for neutrinos at 1~PeV~\cite{Loeb:2006tw}:
\begin{eqnarray}
\frac{d\phisfg}{dE_\nu} & = & \frac{ \fxpp  \times 10^{-18}}{\rm GeV \ cm^2 \ s \ sr} \left(\frac{E_\nu}{100 \, \text{TeV}} \right)^{-2} \nonumber \\
 & & \times \exp \left(-\sqrt{\frac{E_\nu}{1 \, \text{PeV}}} \right)  \, .
\end{eqnarray}
\normalsize
This choice of the spectral shape for SBGs is motivated by the calorimetric limit, implying that the secondary spectrum follows the primary spectrum because the interactions are the dominant process and the particle spectrum inside the SBG is not modified by energy-dependent escape processes in that energy range. The cutoff energy is motivated by the knee of the cosmic ray spectrum, which may be somewhat higher than in our galaxy because of larger magnetic fields; note, however, that the accelerators themselves have to produce particles with high enough energies. Alternative scenarios are \eg\ discussed in \Refs~\cite{Chang:2014sua,Romero:2018mnb,Anchordoqui:2018vji}, such as acceleration in superwinds of SBGs.
Our proposed spectrum is also compatible with the extragalactic diffuse gamma-ray background measured by Fermi \cite{Bechtol:2015uqb}, see \Ref~\cite{Murase:2016gly}, applied to the non-blazar contribution.\footnote{Since $\gamma$-rays from $\pi^0$ decays are produced together with the neutrinos from $\pi^\pm$ decays and these secondaries follow the primary spectrum for $pp$ interactions, the observed extragalactic diffuse gamma-ray background (EGRB) constrains the maximally allowed neutrino flux.
Since most of the EGRB is expected to come from AGN blazars, the constraint on the non-blazar contribution is even somewhat stronger~\cite{Bechtol:2015uqb}. Our obtained SBG contribution will satisfy these limits, see \Ref~\cite{Murase:2016gly}.
}

\subsection{Extragalactic $\boldsymbol{p\gamma}$ neutrinos  ($\boldsymbol{\text{X}_{\text{p}\gamma}}$)}

Neutrinos are produced by $p\gamma$ (or $A \gamma$) interactions in astrophysical environments with high radiation densities. The accelerated primary protons (with a non-thermal spectrum)  collide with photons to produce pions. In such  interactions, the spectrum of the neutrinos depends on both the spectra of the primaries and the target photons.
 
We choose TDEs as our baseline source, since we have a model available~\cite{Biehl:2017hnb} which can not only describe the PeV neutrinos, but also the UHECRs in a self-consistent way. Other candidate classes include  Active Galactic Nuclei (see \eg\ \Refs~\cite{Stecker:1991vm,Nellen:1992dw,Stecker:2013fxa,Murase:2014foa,Rodrigues:2017fmu}) and (low luminosity) Gamma-Ray Bursts (see \eg\ \Refs~\cite{Waxman:1997ti,Dermer:2003zv,Murase:2006mm,Murase:2008mr,Hummer:2011ms,Zhang:2017moz}), which however require more study to draw a fully self-consistent picture. There are ways to discriminate between TDEs and other source classes, \eg, by neutrino point source searches, which we will discuss below. Note again that the UHECR connection requires nuclei in the sources, which means that our TDE example is actually based on $A\gamma$ interactions.

Tidal disruption is the process by which a star is
torn apart by the strong gravitational force of a nearby
massive or supermassive black hole. About half of the
star’s debris remains bound to the black hole, and is ultimately
accreted. TDEs as the sources of extragalactic neutrinos \cite{Wang:2011ip,Wang:2015mmh,Dai:2016gtz,Senno:2016bso,Lunardini:2016xwi,Guepin:2017abw,Biehl:2017hnb} have been recently very actively
discussed in the literature.
In our model we assume that the shape of the neutrino spectrum follows the best fit spectrum of \cite{Biehl:2017hnb}, which we refer to as $(d\phi_{\text{TDE}}^{\text{BF}})/(dE_\nu)$,  with a free normalization:
\begin{equation}
\frac{d\phitde}{dE_\nu}(E_\nu)= \fxpg \frac{d\phi_{\text{TDE}}^{\text{BF}}}{dE_\nu}(E_\nu)
\end{equation}
Note that the parameter $\fxpg$ can be directly related to the normalization $G$ defined in \cite{Biehl:2017hnb}
\begin{equation}
\fxpg = \frac{G}{540} = \frac{\xi_{\text A}}{540} \times \frac{\tilde{R}(0)}{0.1 \ \rm Gpc^{-3} yr^{-1}} \, .
\label{equ:g}
\end{equation}
Here $\xi_{\text A}$ is the baryonic loading (defined as the energy injected as nuclei over the total X-ray energy in the Swift energy range 0.4-13.5 keV) and $\tilde{R}(0)$ is the local apparent rate of jetted TDEs. The reference value chosen for $\tilde{R}(0)$ is the rate of white darf-intermediate mass black hole  disruptions inferred from observations, $\tilde{R}(0) \sim 0.01-0.1 \, \text{Gpc}^{-3} \, \text{yr}^{-1}$. Thus, the result of our combined fit can be immediately interpreted in terms of the TDE baryonic loading $\times$ local apparent rate.

\section{Results}
\label{sec:results}

\begin{figure*}[t]
\centering
\includegraphics[scale=0.45]{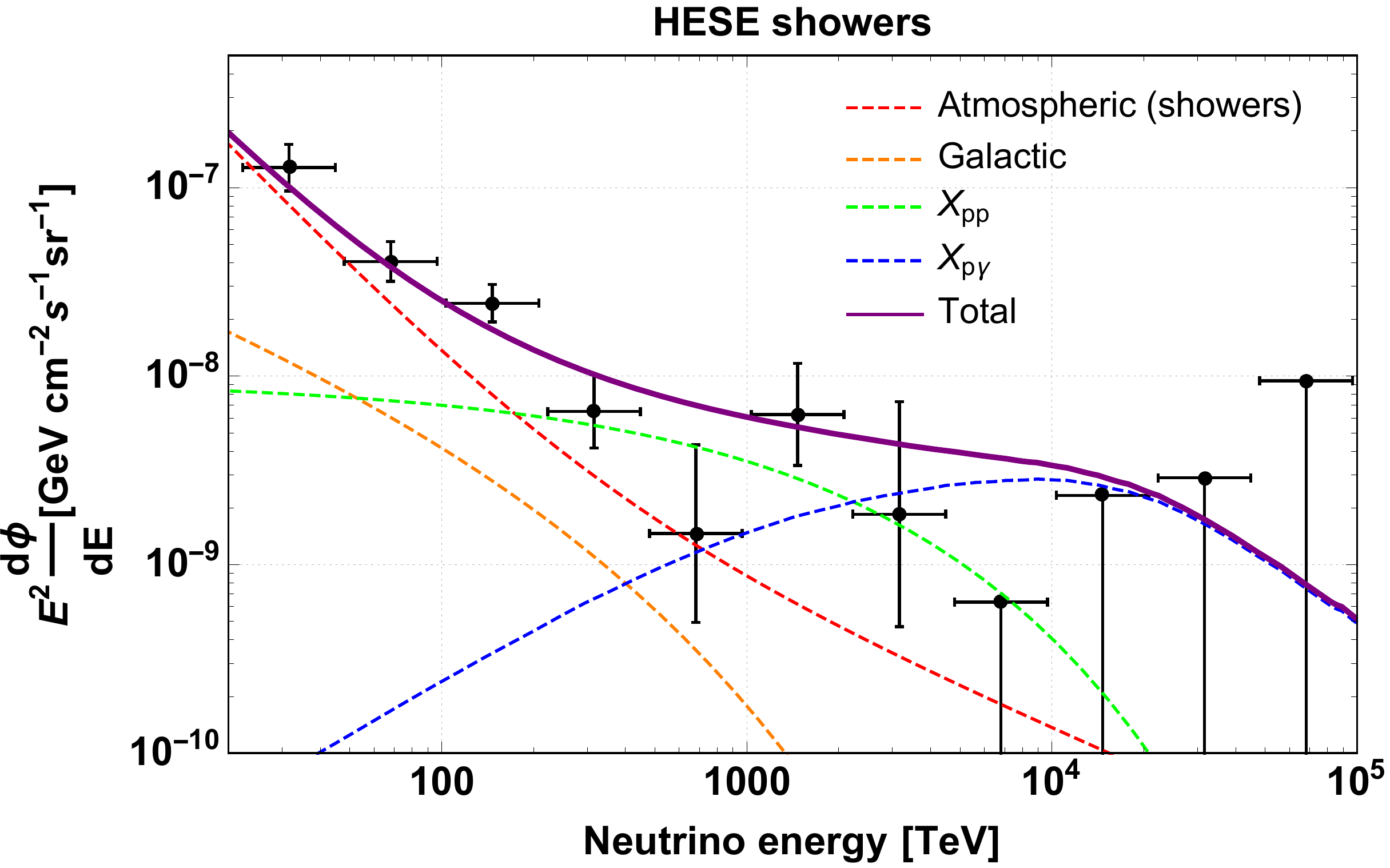} \\[0.5cm]
\includegraphics[scale=0.45]{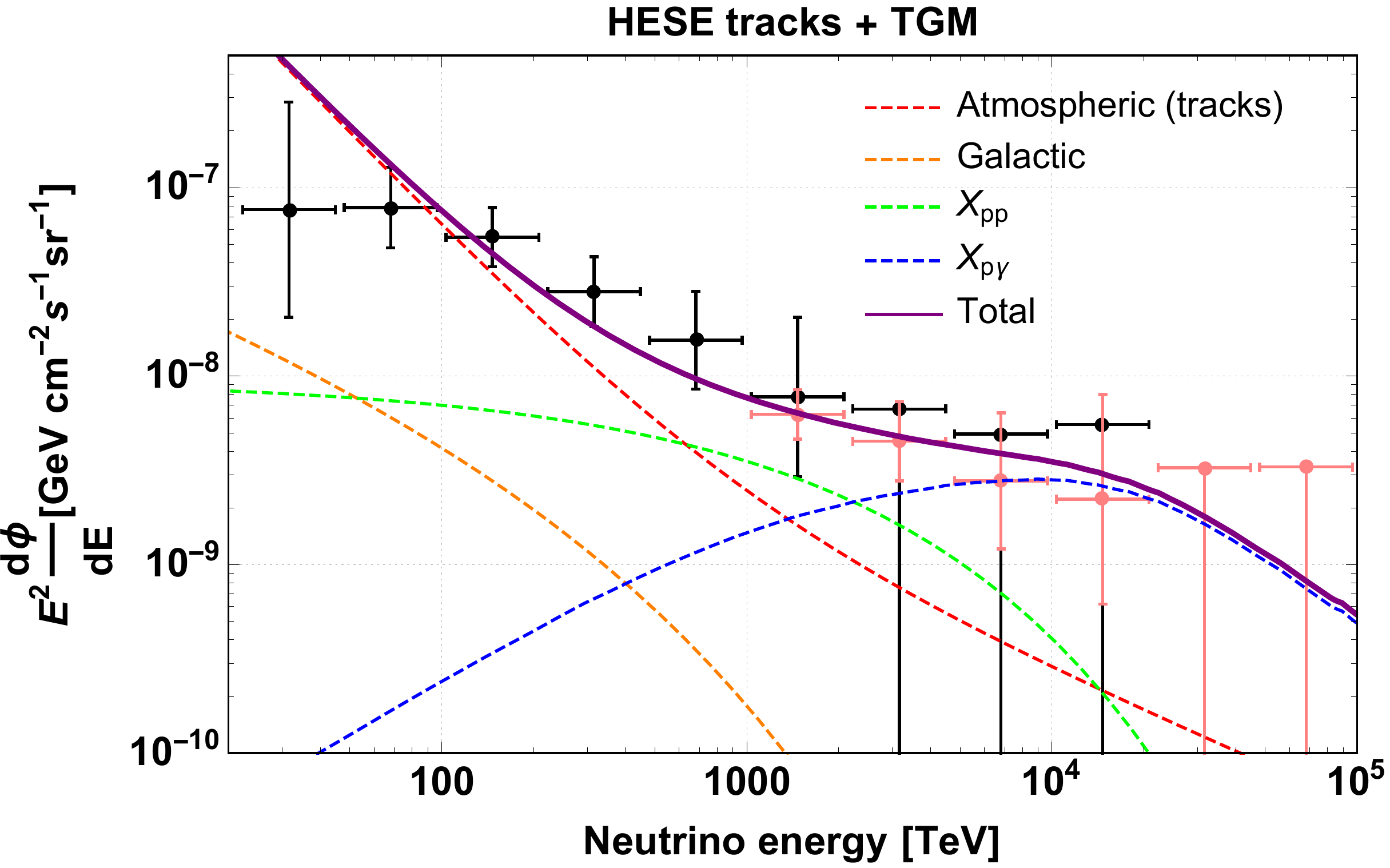}
\caption{Best-fit model (curves) and unfolded IceCube energy spectrum (points with error bars) for showers (upper panel) and tracks (lower panel) using our method described in \App~\ref{app:methods}, where the best-fit for tracks in \Tab~\ref{tab:bf} has been used. Black dots (and error bars) refer to HESE data and pink/light dots (and error) bars to TGM -- which contains the 4.5~PeV reconstructed energy track. The energy scale refers to reconstructed neutrino energy (data) and incident energy (model).  The different contributions (and the total) for the multi-component model are shown, as indicated in the plot legends. Note that, compared to the IceCube analysis, the residual atmospheric background is shown explicitly both in data and model.}
\label{fig:recpoints}
\end{figure*}

\begin{figure*}[t]
\centering
\includegraphics[scale=0.6]{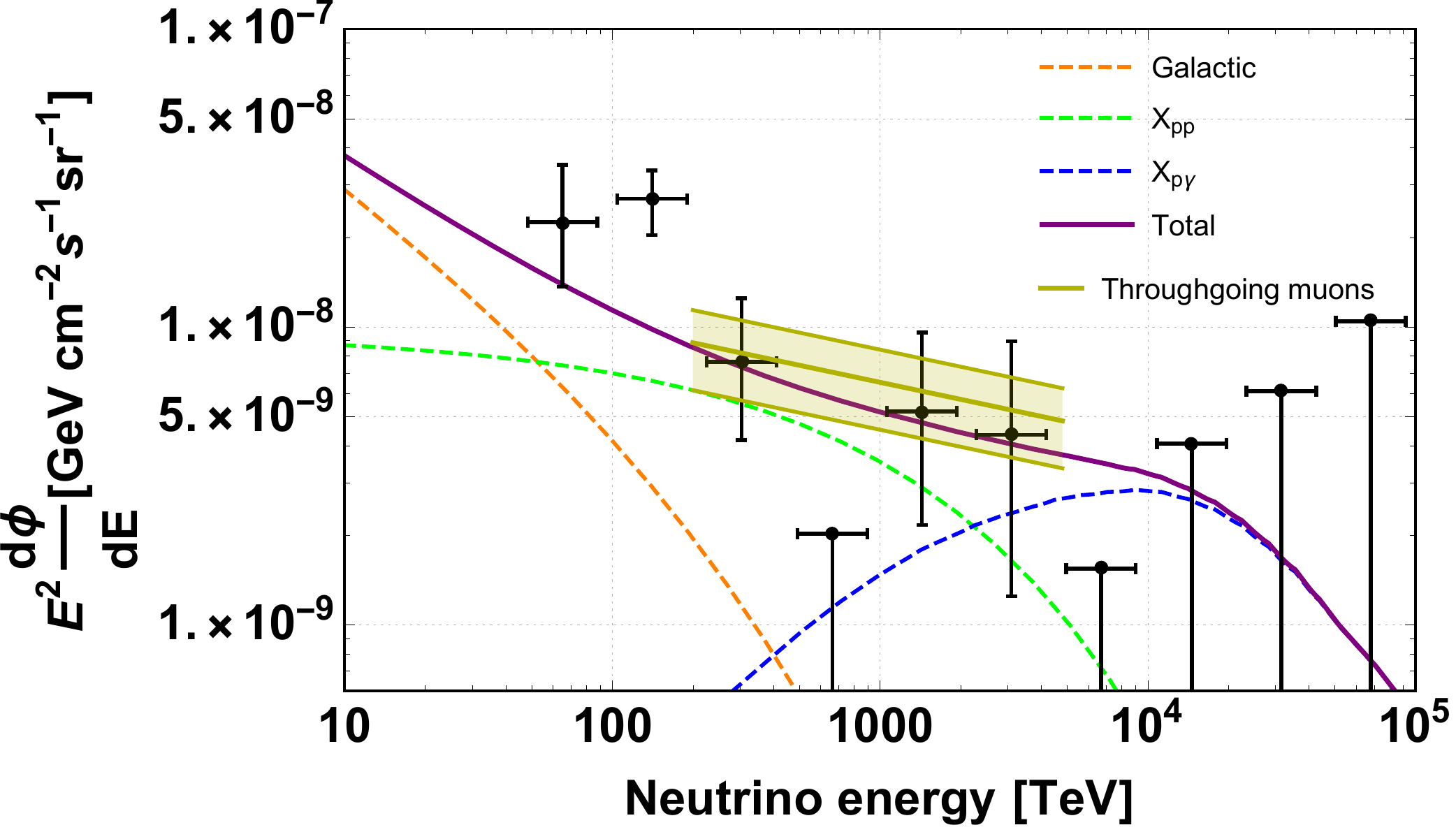}
\caption{Comparison between our multi-component model and the latest IceCube analysis including 6 years of HESE (black points) and 8 years of TGM (yellow band) from \Ref~\cite{Aartsen:2017mau} (neutrino flux per flavor). 
We notice good agreement above a few hundred TeV both with HESE and TGM, whereas the two points at low energies are in tension with the model. Compared to \figu{recpoints}, here the atmospheric backgrounds have been subtracted in both the data points and the model.}
\label{fig:comp}
\end{figure*}

In this section we present our results, including the best-fit model and the result of our energy reconstruction technique. We also demonstrate how to determine the origin of individual events at the probabilistic level.

\subsection{Best-fit model}
\label{sec:bestfit}

We perform a maximum likelihood analysis using a Poissonian $\chi^2$ with the different components introduced in \Sec~\ref{sec:multi-component} and the free parameters $F_{\text{atmo}}^{\mu}$, $F_{\text{atmo}}^{e}$, $F_{\text{gal}}$, $F_{\text{X-pp}}$, and $F_{\text{X-p}\gamma}$. The data points for the different tested data sets are reconstructed in energy as described in \App~\ref{app:methods}, and the theoretical rates are computed with the effective areas similar to \equ{aeff}. 

\begin{table*}[tbh]
\centering
\begin{tabular}{lllllll}
\hline
& $F_{\text{atmo}}^{\mu}$ & $F_{\text{atmo}}^{e}$  & $F_{\text{gal}}$  & $F_{\text{X-pp}}$  & $F_{\text{X-p}\gamma}$ & $\frac{\chi^2}{\text{d.o.f}}$ \\
\hline
Showers (HESE) & & 2.07  &  0.97 & 0.91 & $\sim$ 0 & 1.50 \\
Tracks (HESE+TGM) &  5.73 & & 0.94 & 0.96 & 0.81 & 0.94  \\    
Combined &  5.74& 2.07 & 0.96 & 0.93 & 0.37 & 1.11  \\
\hline
\end{tabular}
\caption{\label{tab:bf} Obtained flux normalizations at the best-fit for showers (HESE), tracks (HESE+TGM), and combined fit (both showers and tracks).}
\end{table*}

\normalsize

Let us first use the shower (HESE) events and the track (HESE + TGM) events separately. 
We obtain the best-fit parameters listed in \Tab~\ref{tab:bf} for the spectral fit in this case. This set of parameters is in agreement with the present constraints: \refmod{The obtained Galactic flux does not violate any present limit, as it can be noticed looking at Fig.3 of \cite{Albert:2017oba}, where different experimental and theoretical limits are illustrated. It is also interesting to notice that the obtained Galactic flux is consistent with the theoretical prediction of 40 years ago reported in \cite{Stecker:1978ah}.}

\refmod{For what concern the extragalactic flux from $pp$ interaction, we have obtained a normalization that is roughly the same of the one reported in Fig.1 of \cite{Murase:2016gly}. Therefore the associated flux of $\gamma$-rays, considering both contributions provided by the direct $\gamma$-ray flux and the electromagnetic cascade, is compatible with the non blazar contribution of the diffuse gamma-ray background, i.e.\ the 20\% of the measured extragalactic diffuse $\gamma$-ray emission. Moreover, we have also checked that the $\gamma$-ray flux associated to our $pp$ neutrinos is compatible with the one reported in Fig.5 of \cite{Tamborra:2014xia}, where the $\gamma$-ray flux has been obtained starting from the infrared radiation of the Starburst Galaxies.}

In addition to, $F_{\text{atmo}}^{\mu}$ and $ F_{\text{atmo}}^{e}$ reproduce the total atmospheric background expected by IceCube. However, they  do not reproduce the expected number of tracks and showers separately. A mis-identification of tracks as showers at the level of 30\%-40\% could reconcile the observations and the expectations. Note again that this problem related to the background is also present in the IceCube analysis itself, since in \cite{Aartsen:2017mau} about 40 tracks are expected (from signal, conventional atmospheric neutrino background and atmospheric muons) -- instead of the 22 tracks detected. 

Most importantly, we find in \Tab~\ref{tab:bf} an almost equal normalization for the Galactic and $\text{X}_{\text{pp}}$ contributions, independent of the sample (tracks or showers used). \refmod{Note that this result only reflects the spectral distribution of the observed events. Their angular distribution and the relevance of the Galactic component in the inner Galaxy will be discussed later (see Sec. \ref{sec:map} and Fig. \ref{fig:inner}).}
The evidence for $\text{X}_{\text{p} \gamma}>0$ depends on the sample used: whereas the best-fit is zero for showers, it is similar to the TDE best-fit in \Ref~\cite{Biehl:2017hnb} for tracks -- which contains the 4.5~PeV reconstructed energy TGM. We use the value obtained from the tracks in the following, as we need to describe both showers and tracks. The combined fit (showers + tracks) yields, instead, $\text{X}_{\text{p} \gamma} \simeq 0.37$, which is non-vanishing but a factor of two smaller than the one suggested by the track sample only.
The $\chi^2/\text{d.o.f.}$ are reported in \Tab~\ref{tab:bf} as well; a value of around one means that that the model describes data very well, while at the same time it is not ``over-fitting'' data.

Our main result is shown in \figu{recpoints}, where we show showers (upper panel) and tracks (lower panel) separately, including the contributions from the individual components. We use the track best-fit here since it is in good agreement with the shower one at low energy, and we expect that it is more representative for our model at high energy since the statistics is much greater above several hundreds of TeV than for the shower sample.
Note that the reconstructed data points are consistent with the IceCube results, and that our figure contains the residual atmospheric background both in data and fluxes explicitly  (see discussion below). 
From the figure it is clear that our model describes both tracks and showers extremely well. At low energies, there are no obvious contradictions with the residual atmospheric backgrounds. At high energies, the possible evidence for $\text{X}_{\text{p} \gamma}>0$ emerges from the TGM dataset (pink), especially the 4.5~PeV track -- which is likely to be generated by an incoming neutrino above 10 PeV. As possible alternative would be to extend $\text{X}_{\text{pp}}$ to higher energies, but this would throw up the question if SBG can accelerate primaries to energies significantly beyond 100~PeV.

We remark again that we show shower and track events separately since they are affected by different backgrounds. Particularly the HESE track sample is more affected by the atmospheric background, since not only atmospheric neutrinos, but also atmospheric muons contribute to the background. For this reason the HESE shower dataset is more representative of the astrophysical signal at low energies, whereas the TGM dataset is more representative of the astrophysical signal at high energies due to the larger effective area related to TGM compared to HESE. 

One of the questions we need to address, is how many components we actually need. We can do that at the statistical level by removing the components one-by-one and re-performing the fit. The atmospheric backgrounds are inevitable as they are needed to the describe the spectrum below 100 TeV; this is also statistically evident. Removing the Galactic component, the quality of the spectral fit is only marginally affected. However,  we have to take into account that the Galactic component also plays a role to explain the angular distribution of HESE events above 100 TeV and the spectral difference between HESE and through-going muons. The presence of a Galactic component, mainly observed from the Southern hemisphere\footnote{70\% of the Galaxy is observed from the Southern sky, whereas only 30\% of it can is observed from the Northern hemisphere (see Fig.~2 of \cite{Palladino:2016zoe}).}, helps to reduce the difference between the observations from South and North. 
Our Galactic component contributes  25\% to the total signal and 30\%-35\% of the astrophysical events coming from South.

The $\boldsymbol{\text{X}_{\text{pp}}}$ component provides the largest contribution to the astrophysical signal, which is about 50\%. The flat $E^{-2}$ spectrum is necessary to explain the middle range of energy between 100 TeV and 1 PeV. Removing this component, the quality of the combined fit goes from  $\chi^2_{\text{min}} = 14.4$ to $27.0$, implying a $3.5 \sigma$ evidence for the $\boldsymbol{\text{X}_{\text{pp}}}$ contribution (even in the presence of 
$\boldsymbol{\text{X}_{\text{p} \gamma}}$ and after re-marginalization). 
Moreover an extragalactic $pp$ spectrum, generated in sources such as SBGs, can explain the lack of correlations between neutrinos and known sources of $\gamma$-rays, since SBGs are rarely observed in $\gamma$-rays because the photons are reprocessed inside the sources. SBGs are also compatible with the point source bounds, as they are faint enough to avoid multiplets from the same source~\cite{Kowalski:2014zda,Ahlers:2014ioa,Murase:2016gly}.

The evidence for the $\boldsymbol{\text{X}_{\text{p} \gamma}}$ contribution is much weaker from the purely statistical perspective only.\footnote{Here $\chi^2_{\text{min}} = 4.7$ goes to $5.7$ using the through-going muon dataset, whereas it remains almost the same using both HESE and TGM, since there is a slight tension between HESE (that disfavor the $p\gamma$ component) and TGM (that favor the  $p\gamma$ component).}
If one used the fit from the track sample,  the $p\gamma$ component would be responsible for about 5-6 events, which are the most energetic ones. Future measurements and particularly IceCube-Gen2, with a 6-8 times greater exposure  \cite{Aartsen:2014njl} can clarify the situation, confirming or excluding the presence of a $p\gamma$ component at PeV energies.  If interpreted as jetted TDE contribution, we find about 80\% of the flux normalization in  \Ref~\cite{Biehl:2017hnb}, which can be interpreted \eg\ in terms of a lower baryonic loading $\simeq 430$, \cf\ \equ{g}. 

\begin{figure}[t]
\centering
\includegraphics[width=0.8\columnwidth]{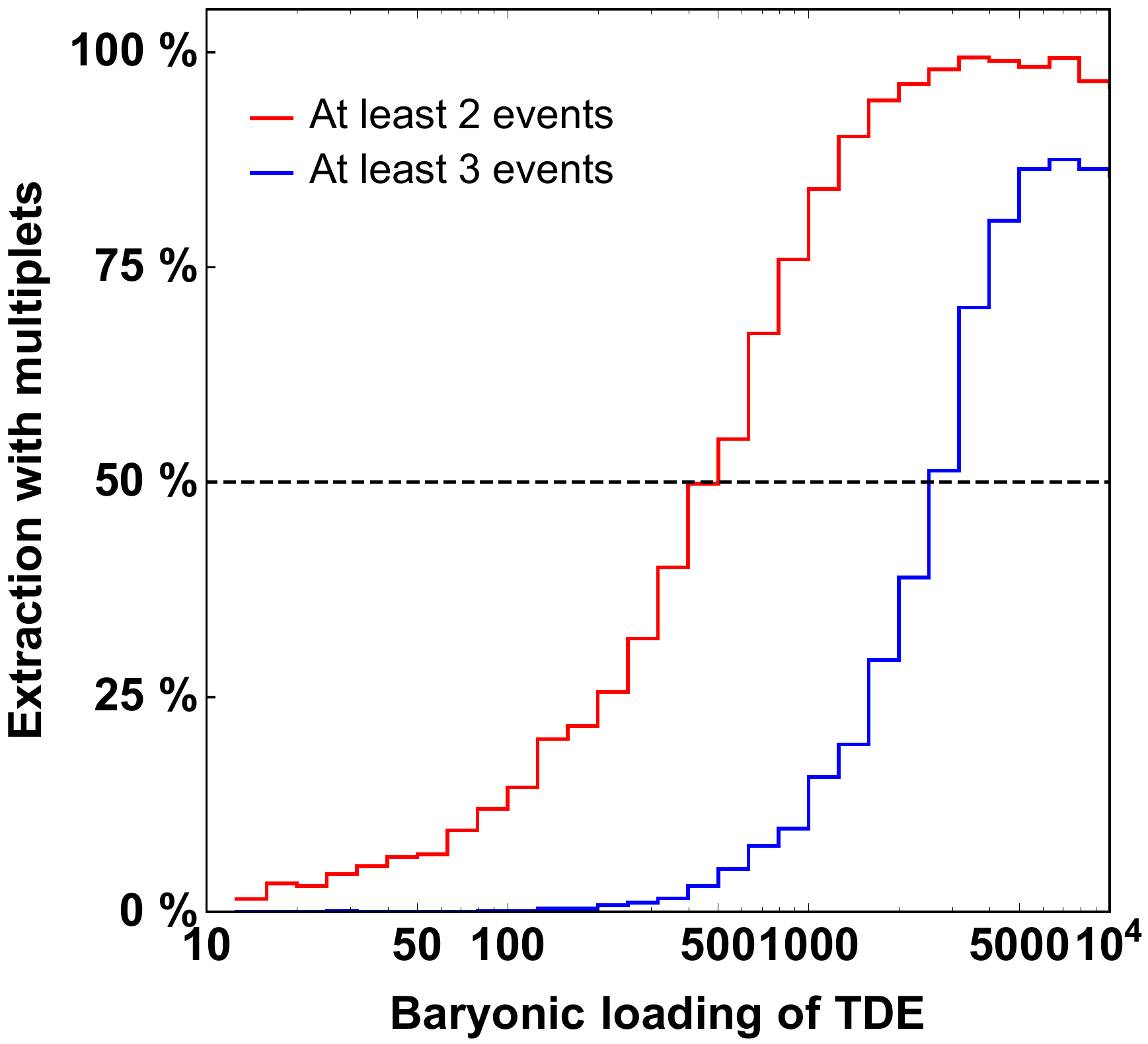}
\caption{Probability to observe multiplets of at least two events (red line) or three events (blue line) from the same TDE as a function of the baryonic loading, considering the present IceCube exposure (5.7 years).}
\label{fig:multipletti}
\end{figure}

Since TDEs are  potentially very luminous neutrino sources, one would eventually expect several events from the same source. We compute the probability to observe at least two or three events from at least one TDE explicitly in \figu{multipletti} as a function of the baryonic loading, assuming that the typical TDE flux corresponds to the one in \cite{Lunardini:2016xwi} (Fig.~2, upper-left panel).\footnote{Using the IceCube HESE effective areas (corresponding to the average TDE at random position), we find that the expected number of events coming from a single TDE $N \simeq  0.24 \times \xi_A/540$. Using our multi-component model and the present IceCube exposure, \ie, 5.7 years for HESE, we find that about six events can be attributed to TDEs. This means that about 25 TDEs contribute to the flux. From this information, \figu{multipletti} can be explicitly computed following \Ref~\cite{Palladino:2017aew} using a Monte Carlo method.}
From this figure, it is evident that the probability to observe two events from one TDE is less than 50\% in the current sample, and the probability for a triplet is at the percent level. 
Note that the limit for individual TDEs, such as 
Swift J1644+57~\cite{Burrows:2011dn}, can be stronger as the effective area depends on declination.
However, this method will be powerful in the future to identify the origin of the $\boldsymbol{\text{X}_{\text{p} \gamma}}$ contribution and to discriminate TDEs from other classes of sources such as LL-GRBs and AGNs. Other method include the test of a possible flavor transition at the highest energies, which is expected for TDEs~\cite{Lunardini:2016xwi} and GRBs (\eg\ \Ref~\cite{Baerwald:2011ee}) due to magnetic field effects on the secondary pions, mesons, and kaons, but not for AGNs.

In \figu{comp}, we compare our model to the data points of the IceCube analysis in which the background has been already subtracted \cite{Aartsen:2017mau}. We notice a good agreement between our model and the IceCube data points in the energy range between 300 TeV and several PeV, and between our reconstructed points (see pink points of \figu{recpoints}) and the IceCube fit of the through-going muons (yellow band of \figu{comp}). On the other hand, at low energies, a certain discrepancy is present: The observation of the excess at 150~TeV by IceCube could be related to an underestimation of the background, since in their data points the background is already subtracted. The gap around 600 TeV only present in the HESE shower dataset in our analysis, whereas it is neither present in our reconstructed track sample, nor in the IceCube TGM analysis. Therefore it could be simply produced by a statistical fluctuation or it could be an hint of the change of the flavor composition at very high energy, although there are not enough data to confirm the last hypothesis.

A special remark concerns the upper limit close to 7 PeV, which is in the Glashow resonance~\cite{Glashow:1960zz} region. It is plausible that the reconstruction of this data point is strongly affected by the production mechanism. In the IceCube analysis, usually  $\phi_{\nu_e} = \phi_{\bar{\nu}_e}$ is assumed; see for example Fig.~7 of \cite{Aartsen:2013jdh}. This assumption applies to $pp$ interactions --  whereas the flux of electron antineutrinos can be suppressed for the $p\gamma$ production of neutrinos~\cite{Barger:2014iua,Palladino:2015vna,Palladino:2015uoa,Biehl:2016psj}, especially in the damped muon regime, in which the electron antineutrino flux at detection is strongly suppressed.
A comparison to the track sample  (lower panel in \figu{recpoints}) is therefore safer here.
We have also checked that the number of Glashow resonant events in the most optimistic case ($pp$ interactions) is roughly 0.5 after 5.7 years of exposure using the best-fit parameters of the model and the approximation for the effective area from  \Ref~\cite{Palladino:2017qda}, which means that at present, our model is compatible with the non-observation of Glashow events. A similar estimate can be performed for double pulse events \cite{Aartsen:2015dlt} for $E_\nu \gtrsim  0.5$ PeV, \ie, the detection of two separate interaction vertices for $\nu_\tau$. Using the effective area in \Ref~\cite{Palladino:2015uoa} and assuming that only $\nu_\tau$ are created at the source (unrealistic, but most conservative scenario), we expect about 0.5 double pulse events after  5.7 years of exposure for our model. Again,  our model is compatible with the non-observation of  double pulses at present.

\subsection{Event-by-event analysis} 
\label{sec:map}

\begin{figure}[t!]
\centering
\includegraphics[width=0.8\columnwidth]{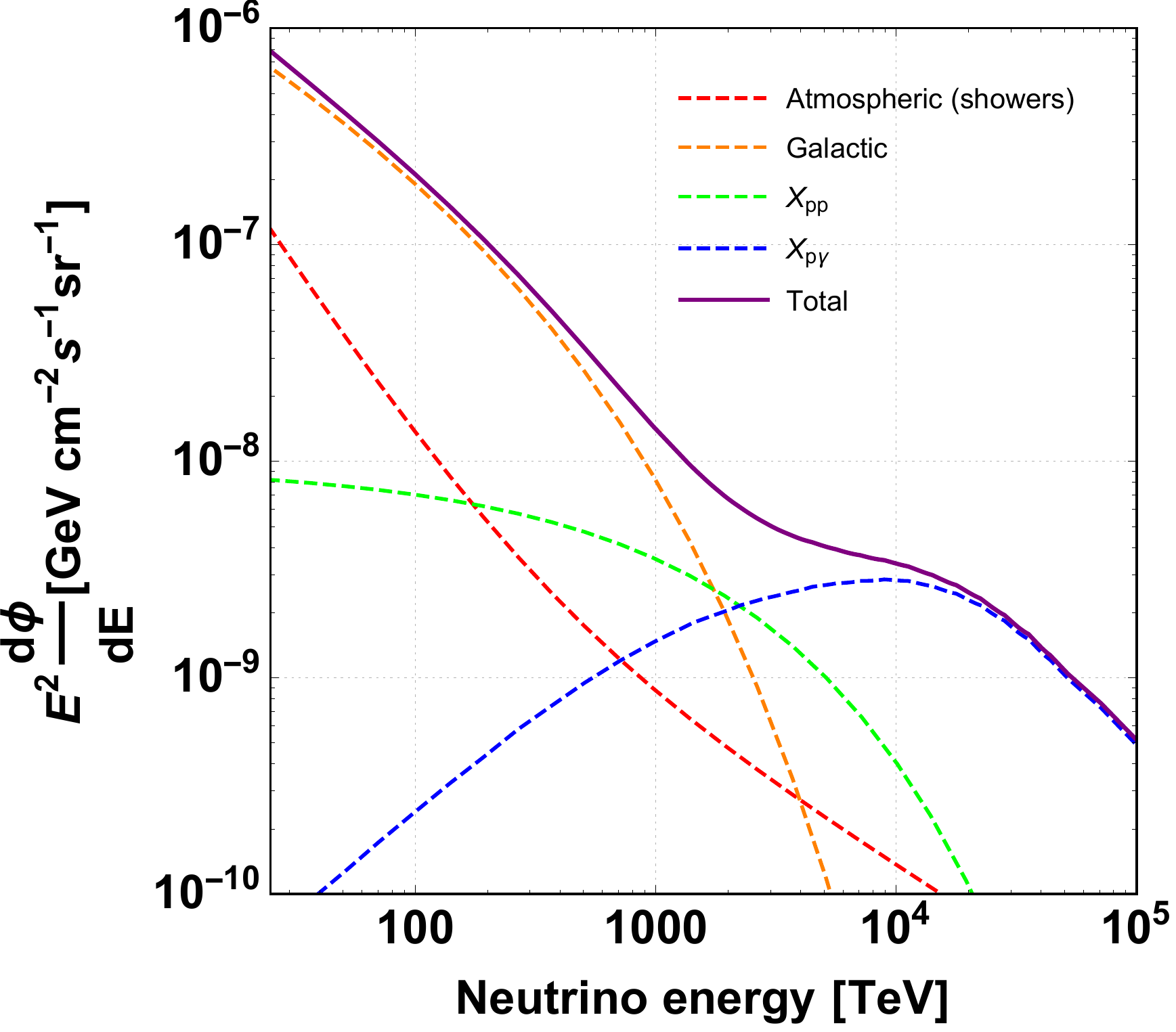}
\caption{\label{fig:inner}
 Fluxes in the inner Galaxy, \ie, the region in Galactic coordinates $|\varphi| \leq 5^\circ$ and $|\lambda| \leq 45^\circ$. The spectra are obtained using the best fit parameters of the multi-component model. The Galactic flux is enhanced in this region compared to the averaged fluxes in \figu{recpoints}, whereas it vanishes outside the Galaxy.}
\end{figure}

\begin{figure}[t!]
\centering
\includegraphics[width=\columnwidth]{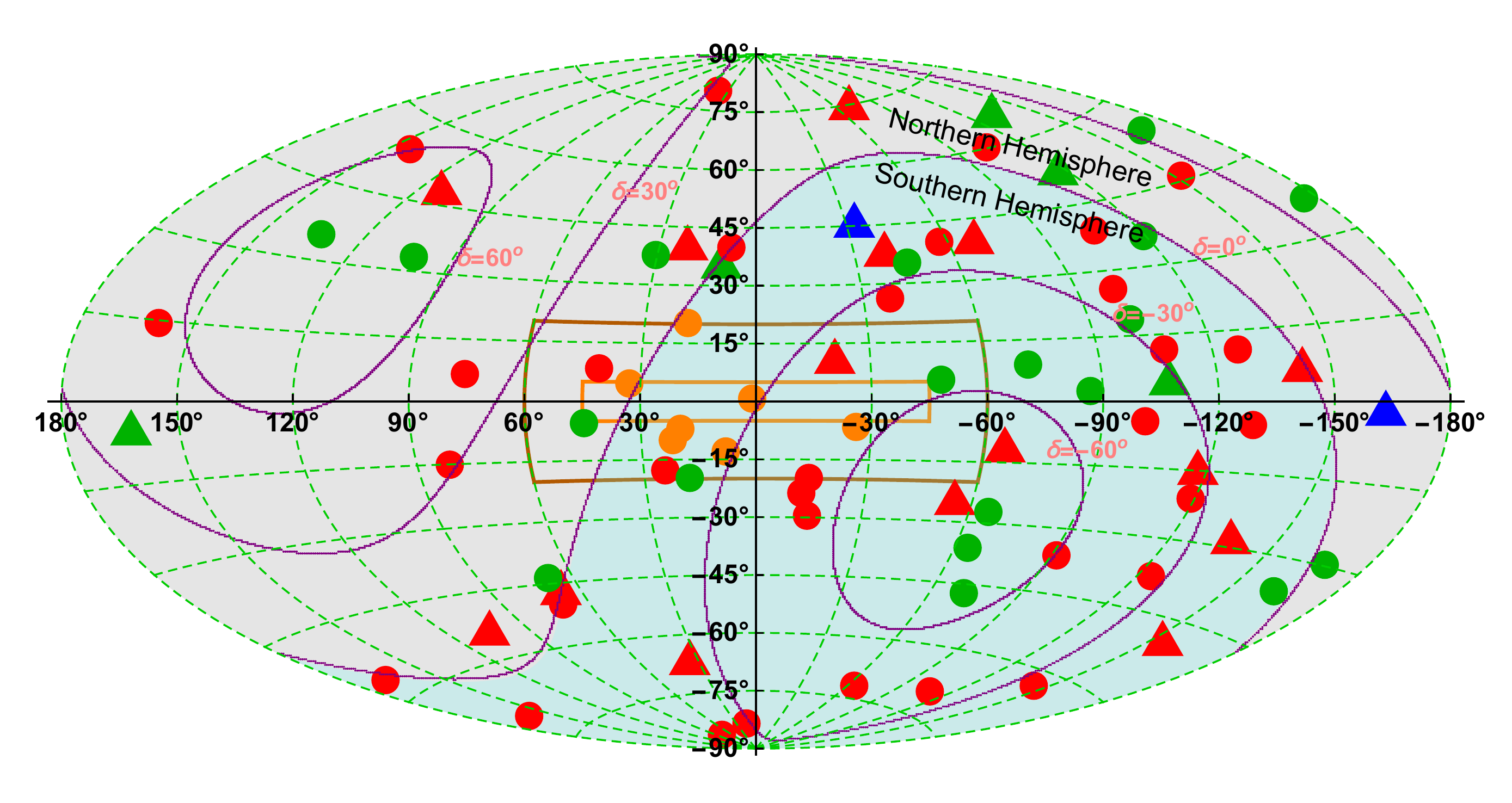} \\
\includegraphics[width=\columnwidth]{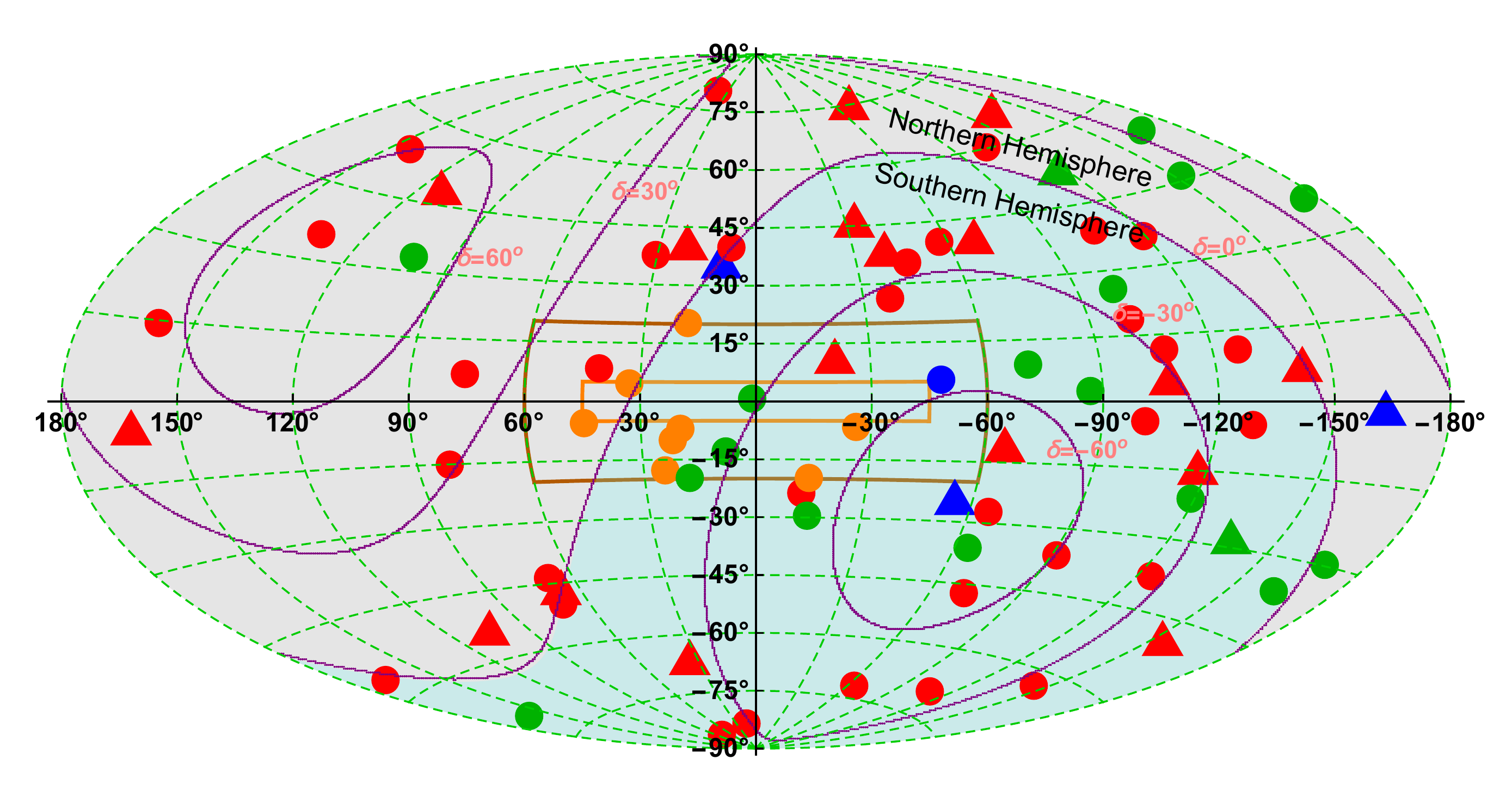} \\
\includegraphics[width=\columnwidth]{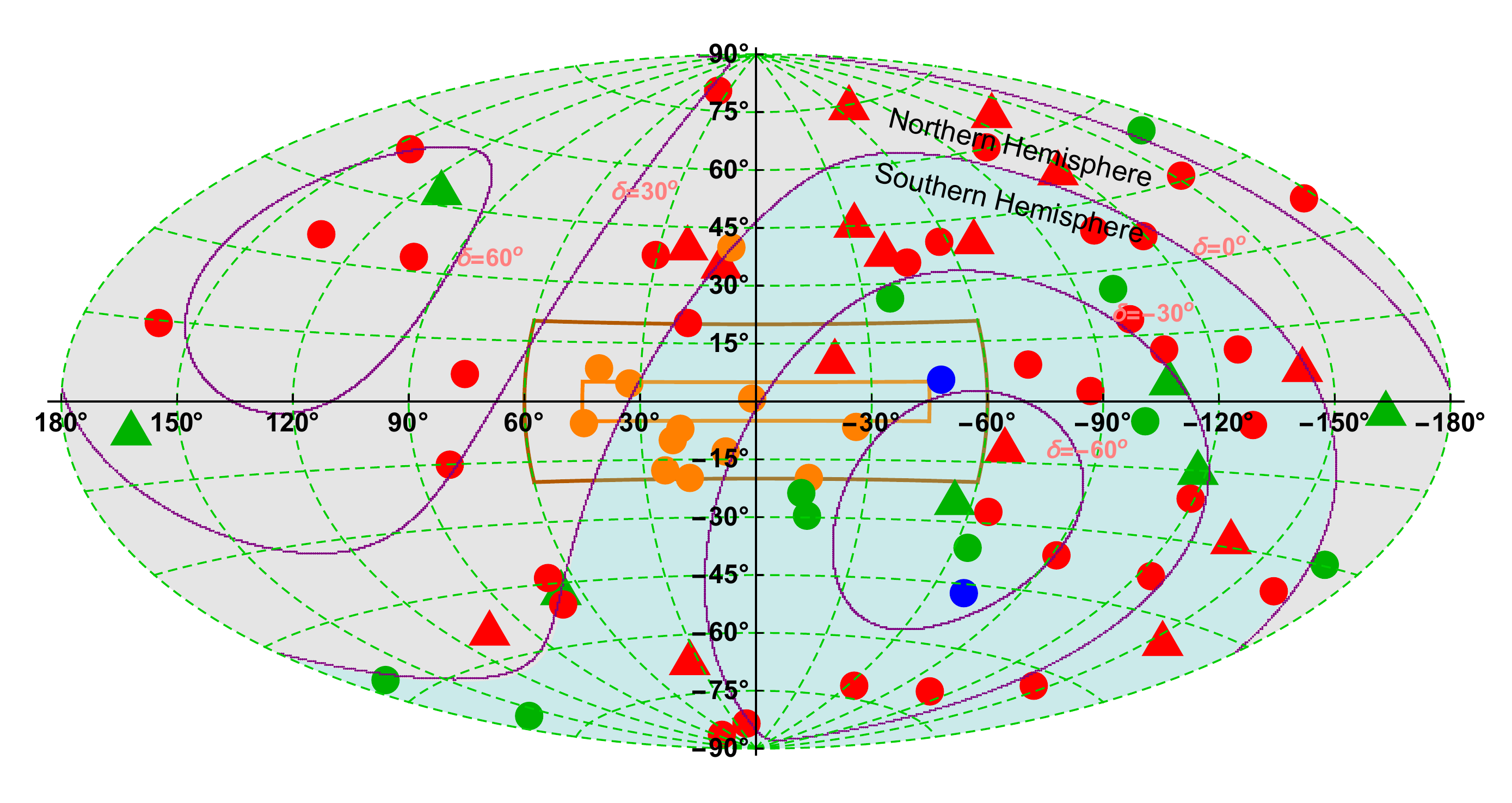}
\caption{Three different examples for the Monte Carlo assignment of the 6 year HESE events to the different components in our multi-component model on an event-by-event basis (Galactic coordinates).  Each color represents a component: red for atmospheric origin, orange for Galactic origin, green for $\boldsymbol{\text{X}_{\text{pp}}}$ and blue for  $\boldsymbol{\text{X}_{\text{p} \gamma}}$. Circles represent showers and triangles represent tracks. The orange lines represent the region in which the Galactic component of high energy neutrinos could be present (light orange) and the region in which a shower-like event is still compatible with Galactic origin (dark orange) within its angular resolution of about $15^\circ$ (on average).}
\label{fig:multimap}
\end{figure}

Using our multi-component model, \refmod{we can compute what is the possible origin of each observed HESE or TGM, based on their reconstructed energy and on their direction.}
This procedure can be sketched as follows (for details, see below):
\begin{itemize}
\item Determine the distribution of the reconstructed energy of each event as described in \App~\ref{app:methods};
\item Determine the probability to belong to each component of the multi-component model (best-fit) according to the spectrum, performing a Monte Carlo extraction of the reconstructed energy; 
\item Determine the probability to belong to the Galactic contribution according to direction and directional resolution.
\end{itemize}

In order to include the direction of each event in our calculation, we need a function related to the compatibility of a certain direction with the Galactic plane, since the Galactic flux is present in the inner Galaxy but absent outside -- including the directional resolution of the events. Note that the Galactic flux actually dominates in the inner Galaxy region up to energies of about 2~PeV, as it is illustrated in \figu{inner}, whereas it is absent outside this region.
Since we have assumed that the Galactic flux is present in the region $|\varphi| \leq 5^\circ$ and $| \lambda | \leq 45^\circ$, we define the directional probability density function such that it is one in that region and drops exponentially outside with the directional resolution:
\begin{eqnarray}
\small
C_{\text{gal}}^h (h) &=& \Theta(m_b -h) + \nonumber \\
& + &  2 \, \Theta (h-m_h) \int\limits_h^{90^\circ} \text{G}(x,m_h,\sigma_h) \ dx \, ,
\label{equ:cgal}
\end{eqnarray}
where G is a Gaussian function similar to \equ{distrox} (but normalized in the range $-\infty$ to $+\infty$) with mean value $m_h$ and standard deviation $\sigma_h$, and $\Theta$ is the Heaviside step function.
We use two functions for Galactic latitude and longitude, \ie, $h$ corresponds to $\varphi$ or $\lambda$; the  product of the previous two functions give the compatibility $C_{\text{gal}} \equiv C_{\text{gal}}^\varphi (\varphi) \times C_{\text{gal}}^\lambda (\lambda)$ of each event with a Galactic origin, based on its direction. Note that we have $m_\varphi=5^\circ$ and  $m_\lambda=45^\circ$ by construction, whereas the standard deviations are different for showers and tracks, for which we use average values  $\sigma_\varphi = \sigma_\lambda \simeq 15^\circ$ and $\sigma_\varphi = \sigma_\lambda \simeq 1^\circ$, respectively (see \eg\ \Ref~\cite{Aartsen:2013jdh}). 
The physical meaning of this function is that an event inside the inner Galaxy is fully compatible with a Galactic origin ($C_{\text{gal}}=1$), whereas the probability drops exponentially to $C_{\text{gal}}=0$ going away from the inner Galaxy. To have the overall probability we have to include the information coming from the energy spectrum.

Therefore the overall probability to belong to a certain component is given by 
\begin{eqnarray}
P_{\ell}(E_\nu) & = & C_{\text{gal}} \frac{\Omega \phi_\ell}{\Omega(\phiatmo+\phisfg+\phitde)+4\pi \phigal} + \nonumber \\
& + & (1-C_{\text{gal}}) \frac{\phi_\ell}{\phiatmo+\phisfg+\phitde}  \, ,
\label{equ:prob}
\end{eqnarray} 
based on the spectrum at the reconstructed energy $E_\nu$. Since the reconstructed energy is not unique, we perform a Monte Carlo simulation (integration) to determine the average probability.

Using this procedure and integrating over $E_\nu$ in \equ{prob}, we can derive the probabilities that an event belongs to the different component. Since the list of events includes the atmospheric backgrounds, it should be clear now why we have carried the residual atmospheric background through the whole procedure. We give the individual probabilities for each event to belong to atmospheric background, Galactic contribution,  $\boldsymbol{\text{X}_{\text{pp}}}$ and $\boldsymbol{\text{X}_{\text{p} \gamma}}$
in \App~\ref{app:prob}. We find for the HESE showers very high probabilities to belong to the atmospheric background in most cases, whereas in eight cases the probability to belong to the Galactic component is higher than 50\%. In 25 cases, the probability to belong to  $\boldsymbol{\text{X}_{\text{pp}}}$ is higher than 30\% (up to around 50\%), whereas only two events exhibit some evidence to belong to $\boldsymbol{\text{X}_{\text{p} \gamma}}$ (probabilities 34\% and 40\%). 
For the HESE tracks, most events seem to belong to the atmospheric background with five exceptions with a significant probability (larger than 30\%) to belong to $\boldsymbol{\text{X}_{\text{pp}}}$. The TGM sample, on the other hand, seems to be dominated by $\boldsymbol{\text{X}_{\text{pp}}}$, whereas one event may be of Galactic origin and 12 events have a probability higher than 30\% to belong to $\boldsymbol{\text{X}_{\text{p} \gamma}}$ -- the 4.5 PeV muon track belong to that component with a 83\% probability. The TGM sample is therefore the sample least dominated by the background and maybe best suited to study the extragalactic neutrino fluxes, whereas the Galactic contribution hides in the HESE showers. \refmod{Our result is comparable with the one obtained in \cite{referee1,referee2}, where the signal is mostly attributed to neutrinos produced in $pp$ and $p\gamma$ sources, identified explicitly as Starburst Galaxies and Blazars in that case. In addition to this, they also found an evidence for an unassociated $E^{-2.7}$ component at low energy, that could be interpreted in terms of residual atmospheric background plus Galactic component in our model.}

This result can be also pictorially represented, see \figu{multimap}, where we show three different samples of the multi-component distribution produced with our Monte Carlo method based on the probability distribution for each event. The different colors corresponds to the different components as described in the figure caption. \refmod{Moreover an information on the topology of the observed events is reported, since the circles represent showers and the triangles represent tracks.} The inner Galaxy region including the corresponding angular uncertainty for showers is also shown (dark orange lines). 
We notice that the events close to the Galactic plane are likely to be associated to Galactic neutrinos in all the maps (orange), whereas the assignment to the component changes in many cases. Some of the events, however, have a high probability to be of atmospheric origin (red). The extragalactic flux is dominated by $\boldsymbol{\text{X}_{\text{pp}}}$ (green), but the assignment to this component can only be performed on a statistical basis. The  $\boldsymbol{\text{X}_{\text{p}\gamma}}$ contribution (blue) tends to apply to the highest energetic events, about 2-3 events in the shown representatives.

\begin{figure*}
\includegraphics[scale=0.35]{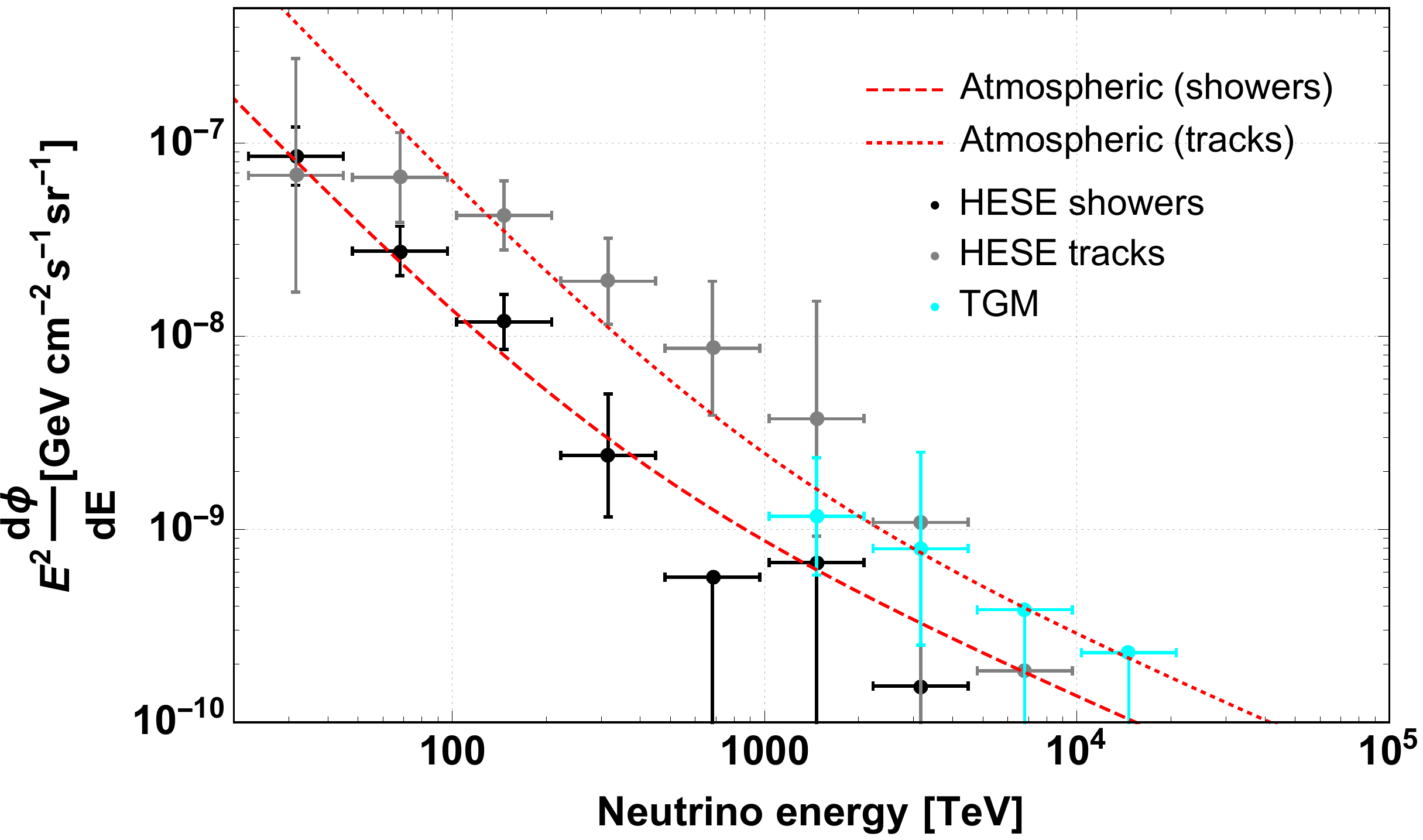} 
\includegraphics[scale=0.35]{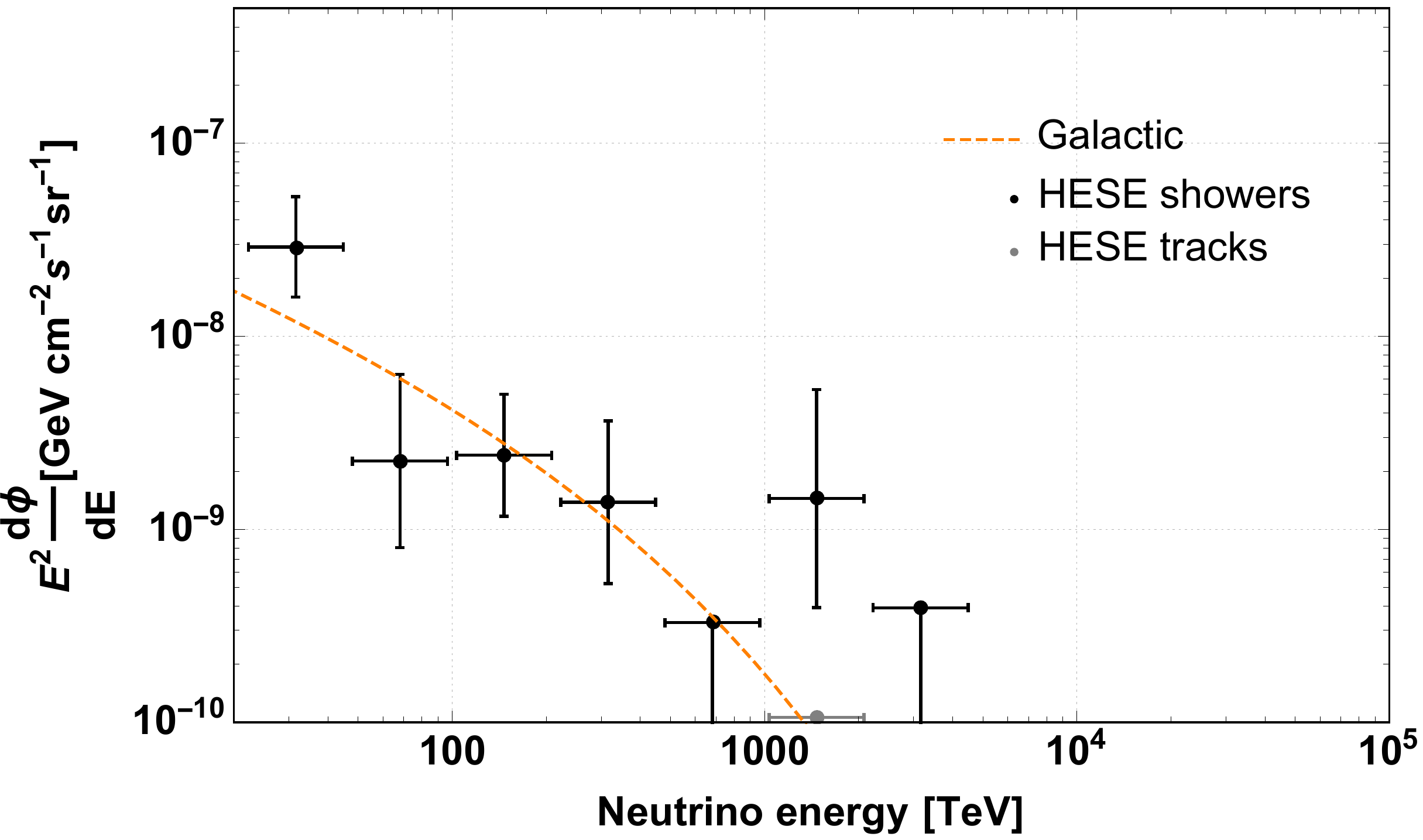}  \\
\includegraphics[scale=0.35]{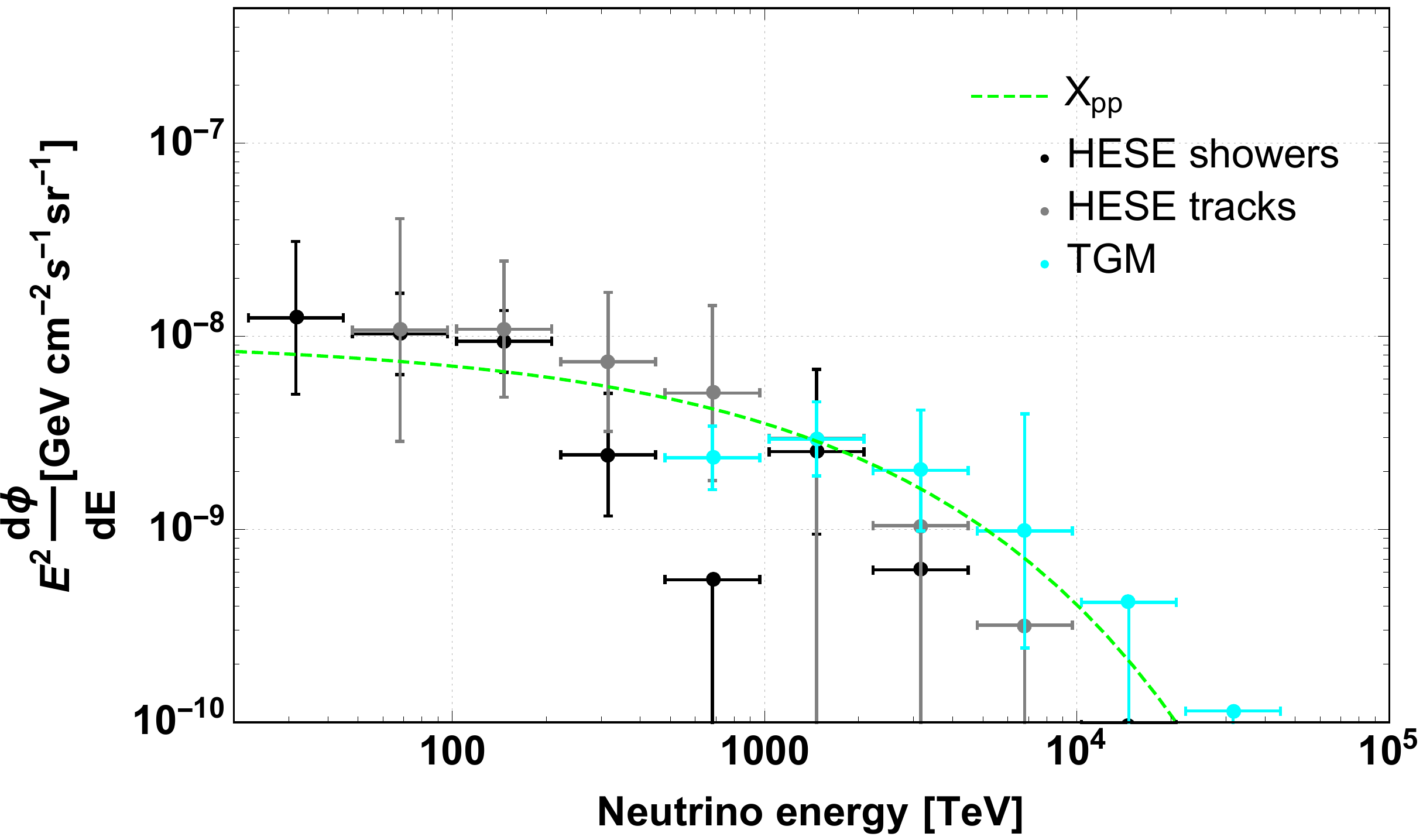} 
\includegraphics[scale=0.35]{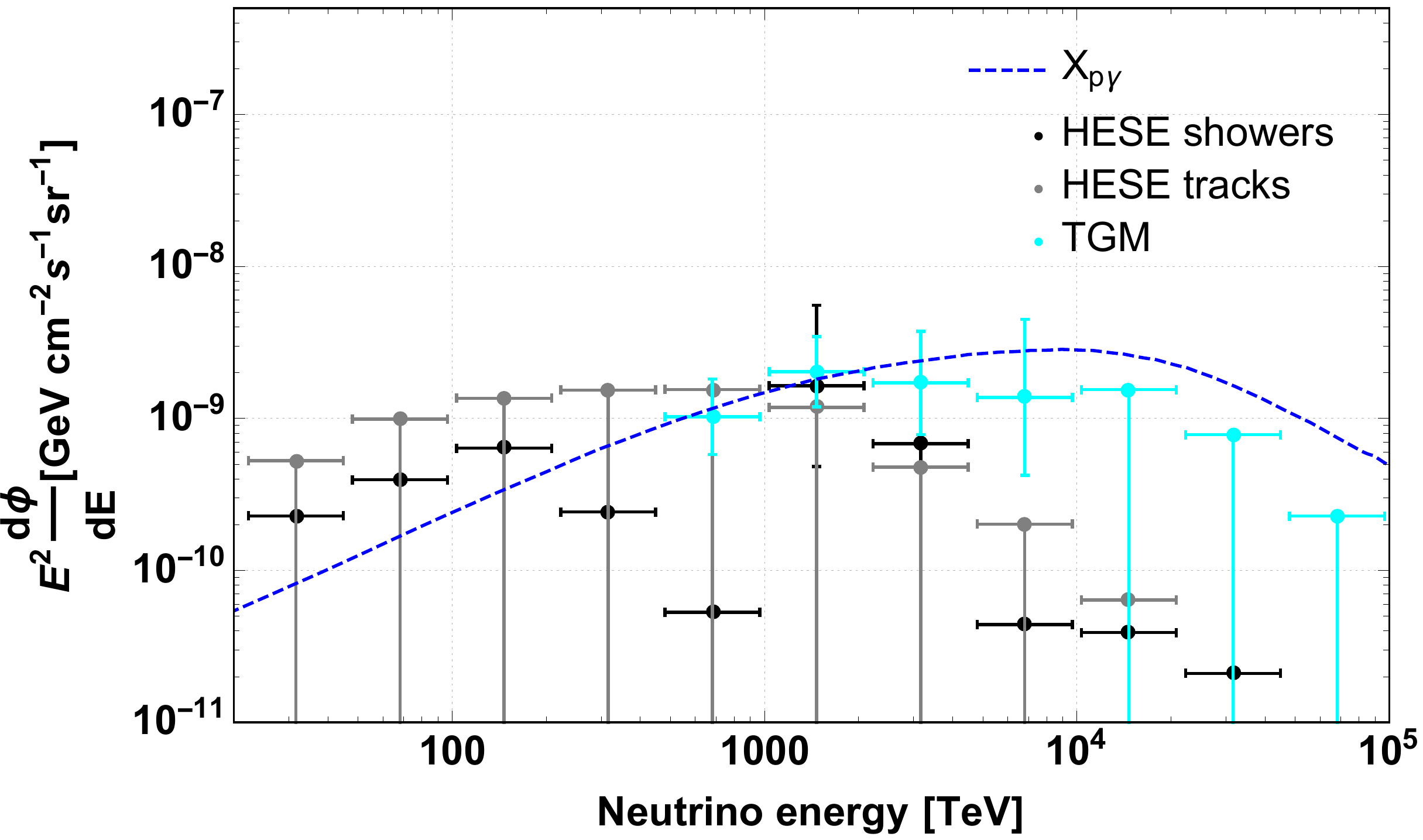} 
\caption{Best-fit model (curves, for track best-fit in \Tab~\ref{tab:bf}) and unfolded IceCube energy spectrum (points with error bars) for HESE showers, HESE tracks and through-going muons using our method described in \App~\ref{app:methods}. Here the data points and error bars are shown for the individual contributions given by the different components, \ie, they depend on the event splitting as given by the fit of the multi-component model.}
\label{fig:4points}
\end{figure*}

We furthermore show in \figu{4points} the unfolded IceCube energy spectrum (points with error bars) for the three samples (HESE showers, HESE tracks and TGM) separated into the different components. Here the  method described in \App~\ref{app:methods} have been used, separating the event rate contributions of the different components.  Let us remark that the data points and error bars are obtained using the probability \equ{prob}, which means that  they depend on our model, whereas  the points reported in \figu{recpoints} are model independent. We can use the figure to see what datasets at what energies describe the different components. For example, one can see that  the residual atmospheric backgrounds are well determined at low energies from HESE data. The HESE showers  constrain the Galactic contribution well around 100~TeV, whereas the $\boldsymbol{\text{X}_{\text{pp}}}$ contribution is equally well contributed to  by HESE showers, tracks, and TGMs. The evidence for $\boldsymbol{\text{X}_{\text{p} \gamma}}$ is mostly driven by the TGM sample.
 
Of course, our analysis is only useful on a statistical basis, but it can be easily extended to future datasets. It allows one to use the information about an individual component if a theoretical model for one source class is studied. There are some limitations of our method. First of all, we do not take into account individual directional uncertainties for the events, which could be used to make the analysis more precise. In order to obtain the uncertainties in Galactic coordinates, a Monte Carlo extraction would be required since the direction and the uncertainties of each event are given in equatorial coordinates and the transformation is not trivial. For the sake of simplicity, we use in \equ{cgal} the average values of the angular resolution  $1^\circ$ and $15^\circ$ for tracks and showers, respectively.

Second, the effective area is actually declination-dependent, whereas we use averaged effective areas. This describes  the isotropic extragalactic component, and it is a good approximation for the Galactic component (as discussed in \Sec~\ref{sec:multi-component}). For the atmospheric flux, we fit the residual atmospheric background passing the vetos and analysis chain with a free normalization independent for tracks and showers. While the effective area for individual events may exhibit a non-trivial declination-dependence, this dependence enters our free normalization as an effective average as long as the residual backgrounds roughly follow the atmospheric flux shape. The deviation between track event rate and residual background  at low energies in \figu{recpoints}, lower panel, may be an artifact such an energy-dependent effect.

\section{Summary and conclusions}
\label{sec:conc}

In this work we have proposed a multi-component model to address the following challenges in current data of astrophysical neutrinos:
\begin{enumerate}
 \item The observed through-going muon spectrum, coming from the Northern hemisphere, is considerably harder than the HESE sample.
 \item
  The HESE dataset exhibits an anisotropy if only events above 100~TeV are considered, indicating a correlation with the Galactic plane.
 \item
  No correlations with known sources have been observed; no point sources have been identified.
 \item
  The limit from the observed extragalactic gamma-ray background must be obeyed.
 \item 
  The recent observation of a muon track with 4.5 PeV reconstructed muon energy requires a primary energy $\gtrsim$ 100~PeV.
\end{enumerate}
We have started from the event topology and deposited energy, and we have computed the probability distribution in terms of event type and reconstructed neutrino energy event-by-event. Consequently, we have reproduced the reconstructed energy distributions by IceCube very well. In comparison to the IceCube publications, we have included the residual atmospheric background explicitly, because we have (at the end) listed the probability for each individual event to belong to each of these different components. Using this approach, we observe two differences compared to the IceCube reconstruction: we do not observe an excess at about 150~TeV, and the gap at about 600~TeV is only present in the shower, but not in the track samples, which points towards a statistical fluctuation (compare data points in \figu{comp} to \figu{recpoints}).

Our model consists of four populations summarized in \Tab~\ref{tab:summarize}: (residual) atmospheric contribution (including prompt contribution) passing the vetos, Galactic contribution following the cosmic-ray flux, extragalactic flux from  $pp$ interactions ($X_{\text{pp}}$), such as starburst galaxies, and extragalactic flux from $A\gamma$ or $p\gamma$ interactions ($X_{\text{p}\gamma}$), such as Tidal Disruption Events or AGN blazars. We have taken the spectra shape for each of these components from plausible examples from the literature and we have fit the (independent) normalizations. We have demonstrated that the above challenges are solved in this model in the following way:
\begin{itemize}
\item 
The through-going muon (TGM) sample is dominated by $X_{\text{pp}}$ and $X_{\text{p}\gamma}$; as a consequence, the measured energy spectrum is close to $E_\nu^{-2}$ in the energy range between 200 TeV and 10 PeV. On the other hand the HESE dataset is more sensitive to low energy events, that are much more affected by atmospheric backgrounds and by the Galactic component, especially for events coming from the Southern hemisphere.
\item Above 100 TeV, the atmospheric background is highly suppressed in the shower HESE dataset, whereas the Galactic component is visible and dominates over the other fluxes in the inner Galaxy region, justifying the 2$\sigma$ excess close to the Galactic plane.
\item The dominant astrophysical $X_{\text{pp}}$  contribution, such as from starburst galaxies, implies abundant and low luminous sources which are in consistency with current stacking or point source searches. Because of the relatively hard spectrum, the bounds from the extragalactic diffuse gamma-ray background can be easily avoided. 
\item
The observed 4.5 PeV track-like event, that is likely to be produced by a $\sim 10$ PeV muon neutrino, justifies $X_{\text{p}\gamma}$. \refmod{As a baseline model we have used jetted Tidal Disruption Events, which are compatible both with the origin of UHECRs and the PeV neutrino flux, as discussed in \cite{Biehl:2017hnb}.}
Since the TDE events are rare but luminous, they can be tested by point source and multiplet limits in the future to discriminate this  hypothesis from alternative scenarios such as low luminosity AGN blazars or GRBs. 
\end{itemize} 
We note that the statistical evidence for $X_{\text{pp}}$ is high ($3.5 \sigma$, even if $X_{\text{p}\gamma}$ is present), while that for $X_{\text{p}\gamma}$ is not significant yet from the purely statistical perspective. The role of $X_{\text{p}\gamma}$ could be taken over by the $X_{\text{pp}}$ component if that was able to accelerate primaries to energies larger than about 100~PeV. While earlier theoretical arguments seem to disfavor this for starburst galaxies, there are recent hints for correlations among UHECRs and starburst galaxies at the observational level by the Auger experiment~\cite{Aab:2018chp}.
An increase of exposure, such as by  IceCube-Gen2, will help to test it: since about 6-7 HESE per year are expected from $X_{\text{p}\gamma}$ at PeV energies, the non-observation of these neutrinos would rule out $X_{\text{p}\gamma}$ at 5$\sigma$ after about 3 years.

We conclude that a self-consistent description of the observed diffuse neutrino flux requires multiple contributions to the diffuse astrophysical neutrino flux, and we have drawn a self-consistent picture describing these contributions by their generic characteristics in terms of spectrum, sky distribution and neutrino production mechanism. Our model solves the key challenges in current neutrino data, including multi-messenger constraints. In the future, it may be helpful to obtain neutrino constraints on the individual contributions in addition to the total spectrum to disentangle the contributions from different source classes. Our model will also help theorists to draw a self-consistent picture if only one of the proposed generic contributions is considered. 

\paragraph*{Acknowledgements}
We would like to thank Anatoli Fedynitch for useful discussions.
This work has been supported by the European Research Council (ERC) under the European Union’s Horizon 2020 research and innovation programme (Grant No. 646623). 

\bibliographystyle{apsrev4-1}
\bibliography{bibliografia}

\appendix

\section{Methods (details)}
\label{app:methods}

In this Appendix,  we describe how we go from the deposited energy per event and the event topology to the reconstructed neutrino spectrum.

\begin{figure*}[t]
\centering
\includegraphics[scale=0.45]{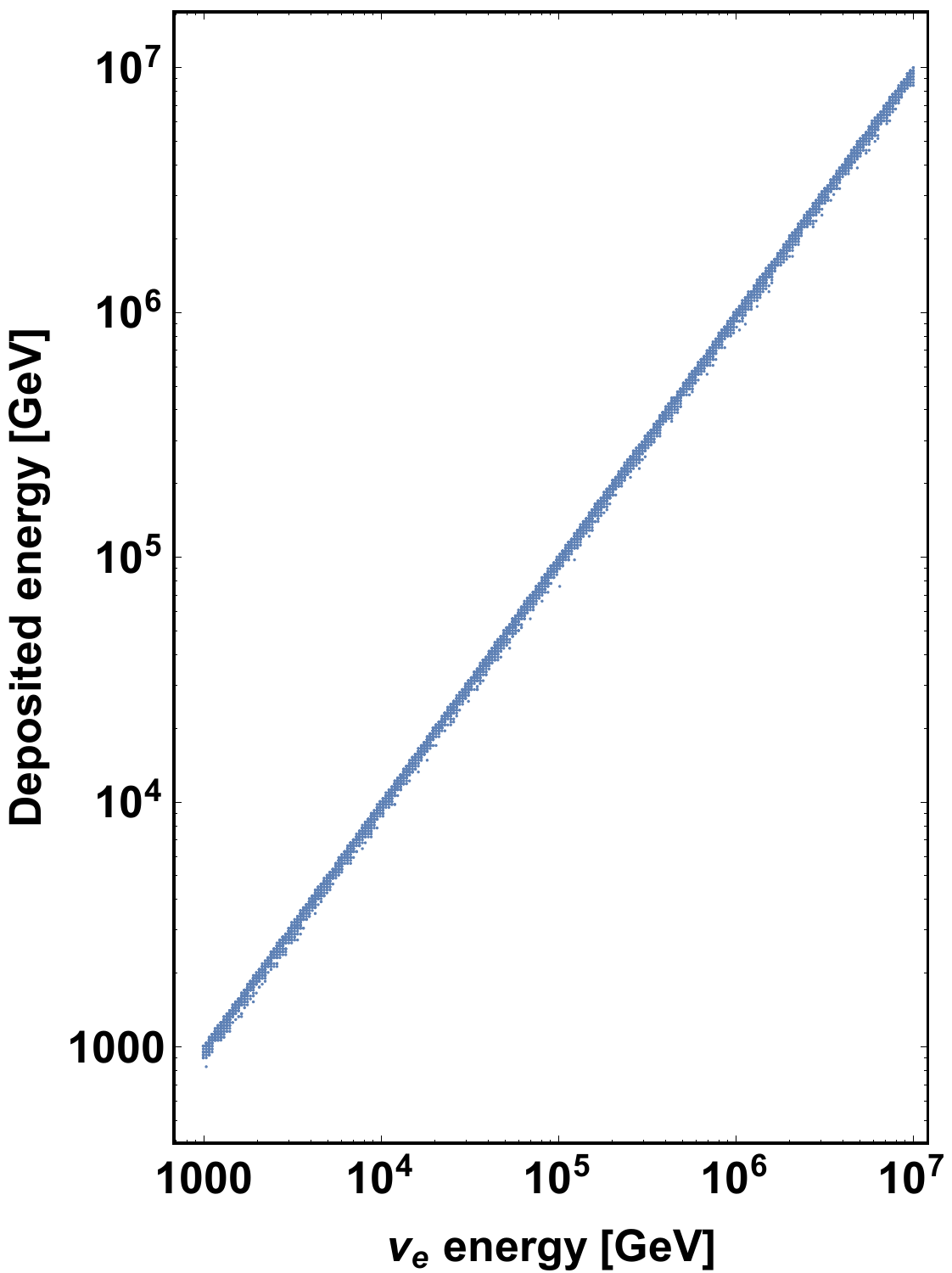}
\includegraphics[scale=0.45]{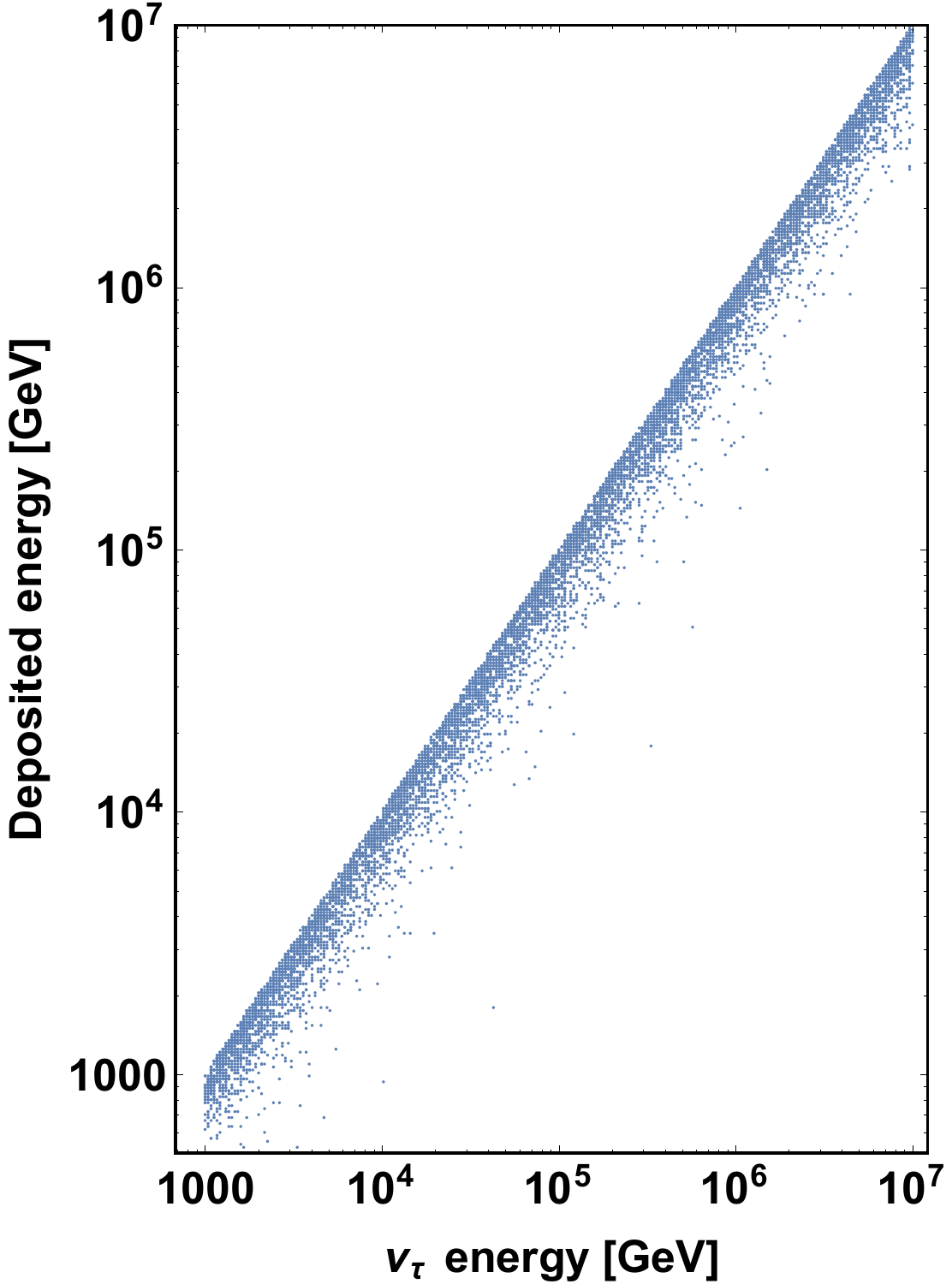}
\includegraphics[scale=0.45]{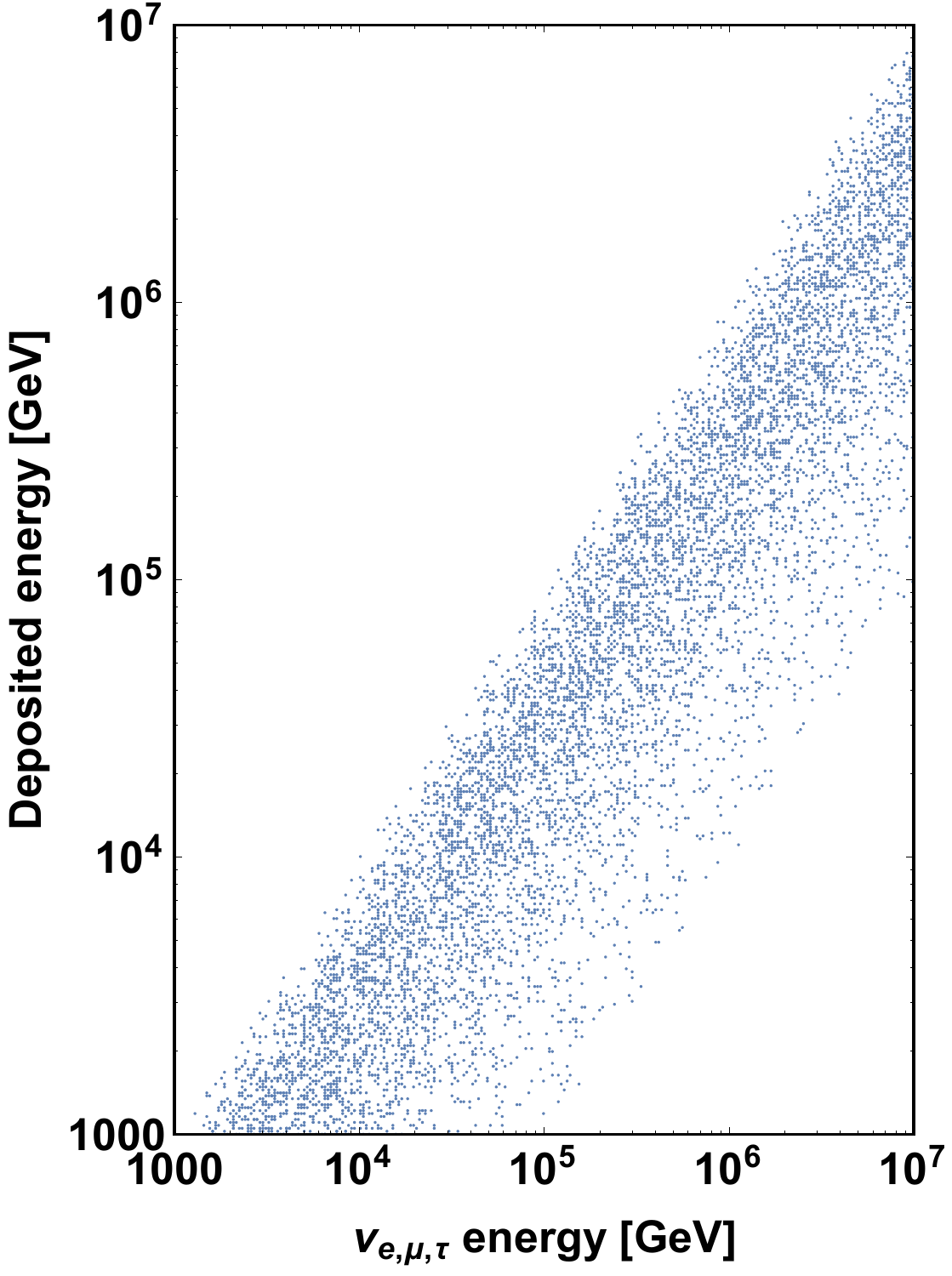}
\caption{Relations between incident neutrino energy and deposited energy   for $\nu_e$ and $\nu_\tau$ CC interactions (left and middle panels)) and for NC interactions (right panel).}
\label{fig:depen}
\end{figure*}

\subsection{Description of the interactions}
\label{app:montecarlo}

Here we define the probability density functions the connect the incident neutrino energy with the deposited energy.  

{\bf $\boldsymbol{\nu_e}$ charged current (CC) interactions.}  
If electron neutrinos (antineutrinos) interact via a charge current interaction, almost the entire energy is deposited into the detector as an electromagnetic shower. Assuming that the fraction $x \equiv E_{\text{dep}}/E_\nu$ of the  incident neutrino energy $E_\nu$ is deposited in the detector, \ie, the deposited energy $E_{\text{dep}}= x\, E_\nu$ with $0 \le x \le 1$, we define a  probability distribution function $\mbox{PDF}_{\text{sh}}^e(x)$ proportional to a Gaussian
\begin{equation}
\text{PDF} \sim \exp \left[ -\frac{(x-m)^2}{2 \sigma^2 }  \right]
\label{equ:distrox}
\end{equation} 
with the normalization
\begin{equation}
\int_0^1 \mbox{PDF}(x) \, dx = 1 \, .
\label{equ:pdfnorm}
\end{equation}
For $\nu_e$ CC interactions, we  use $m_{e,\text{CC}}=0.95$, implying that most of the energy is deposited in the detector,
and $\sigma_{e,\text{CC}}= 0.06$ (see Tab.~1 and Fig.~1 of \cite{Aartsen:2013vja}). 
Our assumption can be compared to the IceCube results comparing the left panel of \figu{depen} with the left panel of Fig.~1 in \cite{Aartsen:2013vja}. 

The probability that a neutrino is a $\nu_e$ and interacts via CC interaction producing a shower-like event is given by
$$
P_{\text{sh}}^e=\frac{\hat \sigma_{\text{CC}}}{\hat \sigma_{\text{CC}}+\hat \sigma_{\text{NC}}} \times \frac{\phi_e}{\phi_e+\phi_\mu+\phi_\tau} \, ,
$$
evaluated at the incident neutrino energy, where $\hat \sigma$ is the cross section. 
The ratio $\hat \sigma_{\text{CC}}/(\hat \sigma_{\text{CC}}+\hat \sigma_{\text{NC}})$ is in good approximation constant in the energy range between 100 TeV and 10 PeV, \ie, our range of interest, and it is equal to 0.71 \cite{Gandhi:1998ri}. Moreover we assume the equipartition of flavors in order to simplify the calculation, which leads to  $P_{\text{sh}}^e \simeq 1/3 \times 0.71 \simeq 23.7\%$.\footnote{Since we will separate tracks and showers  and we will use the relative probabilities for the two different classes of events, this procedure also works correctly also with atmospheric neutrinos.}

{\bf $\boldsymbol{\nu_\tau}$ charged current (CC) interactions.} 
\label{app:nutaucc}
If tau neutrinos (antineutrinos) interact via charged currents, a fraction of  energy is carried away by neutrinos generated by $\tau$ decays which is not detectable. Considering the various channels in which a $\tau$ can decay, the average quantity of energy fraction detectable is about 80\%~\cite{Vissani:2013iga}.  
We also use a Gaussian, see \equ{distrox}, with $m_{\tau,\text{CC}}=1$ and $\sigma_{\tau, \text{CC}}=0.25$ in order to reproduce the average value of the transferred energy. Therefore the uncertainty is larger here than for $\nu_e$ CC interactions, see middle panel of \figu{depen}.   

Moreover $\nu_\tau$ can produce muons with a branching ratio of $\mbox{B.R.}(\tau \rightarrow \mu)=17.4\%$. This contribution has to be subtracted, since we only discuss shower-like events in this section. 
The probability that a neutrino is a $\nu_\tau$ and interacts via charged currents producing a shower-like event is given by:
$$
P_{\text{sh}}^\tau=\frac{\hat \sigma_{\text{CC}}}{\hat \sigma_{\text{CC}}+\hat \sigma_{\text{NC}}} \times \frac{\phi_\tau \, (1-\text{B.R.}(\tau \rightarrow \mu))}{\phi_e+\phi_\mu+\phi_\tau} \, , 
$$
which is $P_{\text{sh}}^\tau \simeq 19.5\%$

{\bf Neutral current (NC) interactions.} 
For NC interactions, which are not sensitive to flavor, a large fraction of the incoming neutrino energy is not visible because it  is carried away by outgoing neutrinos. 
The average fraction of deposited energy is 25\%, but large fluctuation are expected in this case (see Fig.~1, right panel, in  \cite{Aartsen:2013vja}). We therefore use  $m_{\text{NC }}=0$ (events that deposit a small fraction of energy are favored) and standard deviation  $\sigma_{\text{NC}} =0.3$ in \equ{distrox}.
This choice reproduces the average fraction of deposited energy, and it also describes in good approximation the right panel of Fig.~1 in \cite{Aartsen:2013vja} (compare to the right panel of \figu{depen}). 

The probability that a neutrino interacts via neutral currents is given by
$$
P_{\text{sh}}^{\text{NC}}=\frac{\hat \sigma_{\text{NC}}}{\hat \sigma_{\text{CC}}+\hat \sigma_{\text{NC}}} \simeq 29\% \, .
$$

{\bf Track-like events: $\boldsymbol{\nu_\mu}$.}
Track-like events are generated by $\nu_\mu$ that interact via CC interaction and produce a muon, which is visible as track in the detector. For track-like events the information on the incoming neutrino energy is very uncertain and the deposited energy only represents a lower limit to the incoming neutrino energy (the track often originates outside the detector). As for NC interactions the deposited energy is usually a small fraction of the incoming neutrino energy; therefore we use the same probability density function used for NC interaction, \ie, $\text{PDF}_{\text{tr}}^\mu = \text{PDF}_{\text{sh}}^{\text{NC}}$.
We have checked that this approximation reproduces the measured through-going muon spectrum  in good approximation, as it can be seen comparing the pink points of \figu{recpoints} to the yellow band in \figu{comp}. 

The probability that a neutrino is a $\nu_\mu$ and  produces a track-like event is given by
$$
P_{\text{tr}}^{\mu}=\frac{\hat \sigma_{\text{CC}}}{\hat \sigma_{\text{CC}}+\hat \sigma_{\text{NC}}} \times \frac{\phi_\mu}{\phi_e+\phi_\mu+\phi_\tau} \simeq 23.7 \% \, .
$$

{\bf Track-like events: $\boldsymbol{\nu_\tau}$.} Track-like events can be generated, in smaller number, by muons created in the $\nu_\tau$ CC interactions.
The probability that a neutrinos is a $\nu_\tau$ which interacts via CC interaction and produces a track-like event is given by:
$$
P_{\text{tr}}^{\tau}=\frac{\hat \sigma_{\text{CC}}}{\hat \sigma_{\text{CC}}+\hat \sigma_{\text{NC}}} \times \frac{\phi_\tau \ \text{B.R.}(\tau \rightarrow \mu)}{\phi_e+\phi_\mu+\phi_\tau} \simeq 4.1 \% \, . 
$$

Re-call that here the muon is generated by tau decay,
$$
\tau \rightarrow \mu + \bar{\nu}_\mu + \nu_\tau \, ,
$$
which means that the muon takes about 1/3 of the $\nu_\tau$ energy. We therefore use $\text{PDF}_{\text{tr}}^\tau(x) = \text{PDF}_{\text{tr}}^\mu(3 x)$.

\subsection{From deposited to reconstructed energy}

\begin{figure*}[t]
\centering
\includegraphics[scale=0.5]{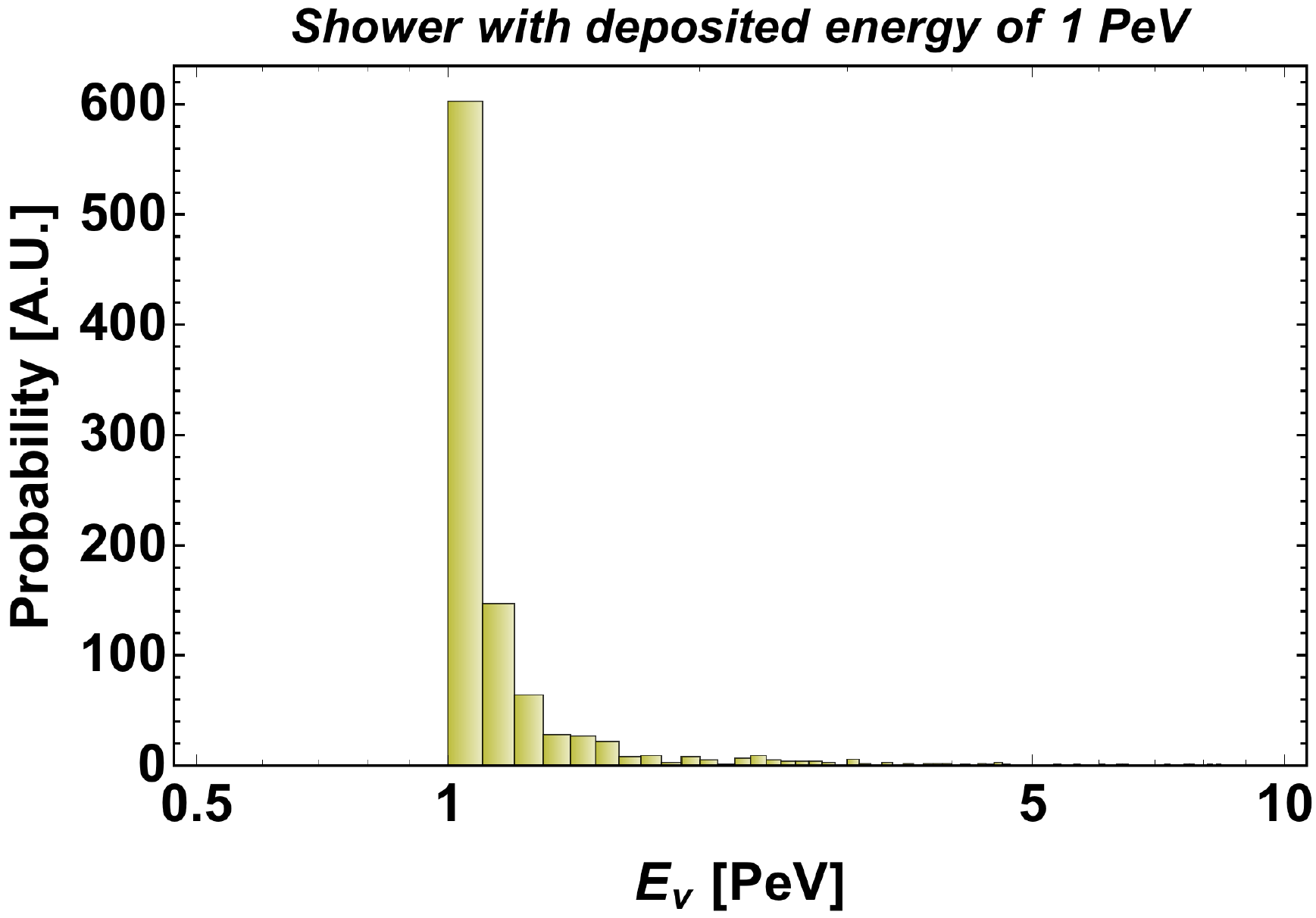}
\includegraphics[scale=0.5]{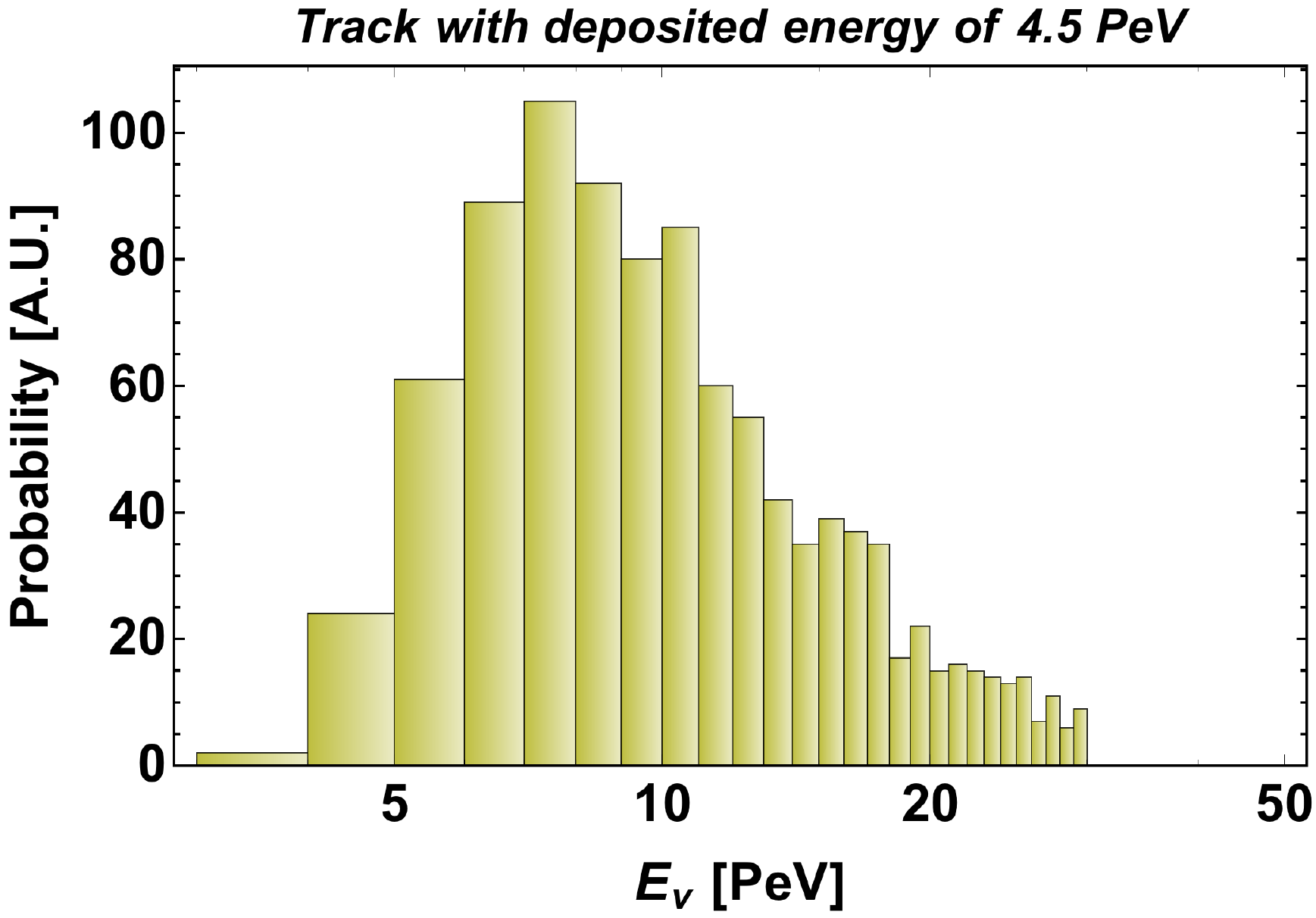}
\caption{Distribution of the reconstructed (neutrino) energy for a shower with deposited energy of 1 PeV (left panel) and a track with reconstructed muon energy of 4.5 PeV (right panel). Here the assumption for the spectral shape is $E^{-2}$. }
\label{fig:en}
\end{figure*}

We know construct the probability density function $R_\ell(E_{\text{dep}},E_\nu)$ (differential in $E_\nu$) of the reconstructed neutrino energy for a given deposited energy and event type $\ell$. Note that for the sake of simplicity we use the symbol $E_\nu$ simultaneously for the incident (true) and reconstructed neutrino energies. 

Compared to the PDFs in the previous section, the function $R_\ell$ will depend on the spectral shape.
We assume an $E^{-2}$ flux for the neutrino spectrum by default, but we  will discuss the consequences of different  assumptions below. Because the  cross section of neutrino-nucleons interaction scales $\propto \sqrt{E}$ in the energy range between 100 TeV and 10 PeV \cite{Gandhi:1998ri}, the function $R_\ell$ is given by
\begin{equation}
R_\ell(E_{\text{dep}},E_\nu) \propto \text{PDF}_\ell\left(\frac{E_{\text{dep}}}{E_\nu} \right) \frac{E_{\text{dep}}}{E_\nu^2} \times E^{-1.5} \, ,
\label{equ:R}
\end{equation}
where the last factor is the product between spectrum and cross section energy dependencies.
Here $\ell$ denotes one of the process previously described, namely showers given by $\nu_e$ CC, $\nu_\tau$ CC or NC interactions and tracks given by $\nu_\mu$ CC and $\nu_\tau$ CC interactions.  The factor $E_{\text{dep}}/E_\nu^2$ comes from the change of variables from $x$ to $E_\nu$ in the normalization \equ{pdfnorm}.
In order to derive this formula, see also Sec.~2 of \Ref~\cite{Huber:2004ka}.\footnote{Our function PDF corresponds to the function $k_f$ therein; replace $T_f \, V_f$ by a $\delta$-distribution as the deposited energy has been measured here.}

The probability that an event with a certain deposited energy has been generated by a specific process becomes
\begin{equation}
\tilde P_\ell =\frac{P_\ell \int R_\ell(E_{\text{dep}},E_\nu) \ dE_\nu }{\sum_\ell P_\ell \int R_\ell(E_{\text{dep}},E_\nu) dE_\nu }
\label{equ:probtilde}
\end{equation}
We have checked that these probabilities, to a very good approximation,
 do not depend upon the deposited energy. The probabilities that a shower-like event has been produced by $\nu_e$ CC, $\nu_\tau$ CC, and NC interactions are 55\%, 35\%, 10\%, respectively. The probabilities that a track-like event has been produced by $\nu_\mu$ CC or $\nu_\tau$ CC interactions are 96\% and 4\%, respectively. 
Note that $\tilde P_\ell$ includes the spectral re-distribution of events, as compared to $P_\ell$, and that we have pulled out $P_\ell$ from the integrals because they hardly depend on energy. For example, $\tilde P_\text{sh}^\text{NC}$ is relatively (compared to the other charged current showers) smaller than  $P_\text{sh}^\text{NC}$ because a NC shower with a certain deposited energy tends to come from a higher energy than a CC shower, where the flux is lower. 

\subsection{Monte Carlo simulation and deconvolved neutrino flux} 
\label{sec:montecarlosub}

Starting from the list of events with event topology (track-like or shower-like) and $E_{\text{dep}}$, we construct the probability distribution for the deposited energy $C_i(E_\nu)$ for the $i$th event as follows:
\begin{enumerate}
\item We determine a process that has generated the observed event from the probability $\tilde{P}_\ell$ in \equ{probtilde}.
\item We extract an energy associated to the process according to the distribution $R_\ell(E_{\text{ dep}},E_\nu)$ in \equ{R}.
\item We repeat this procedure  $10^4$ times to obtain the  reconstructed energy distribution $C_i(E_\nu)$ for this event.
\end{enumerate}
We show in \figu{en} the distribution of the reconstructed (neutrino) energy for a shower with deposited energy of 1 PeV (left panel) and a track with reconstructed energy of 4.5 PeV (right panel), \ie,  the most energetic event observed by IceCube up to now. The reconstructed energy  of a shower-like event is close to the deposited energy and there is a small uncertainty, whereas  the reconstructed energy of a track-like event is two to three times larger than the deposited one, with a much bigger uncertainty.
We have checked that both our probability distribution and our reconstructed energies are in very good agreement with the theoretical estimates in \Tabs~IV and~V of \Ref~\cite{DAmico:2017dwq}, which appeared during the completion of this work. The choice of a softer spectrum instead of the assumed $E^{-2}$ would reduce the uncertainty on the reconstructed energy, privileging the events that deposit most of their energy. We have verified that using an $E^{-2.5}$ spectrum, the uncertainty on the reconstructed energy is reduced 
by about 30\% for shower-like events and by about 25\% for the track-like events, deteriorating both the agreement with \cite{DAmico:2017dwq} and the TGM analysis performed by IceCube~\cite{Aartsen:2015rwa,Aartsen:2017mau}. 

We can now determine the total number of events in each reconstructed energy bin $j$ covering the energy interval $[E_j,E_{j+1}]$, where we choice of the binning is arbitrary. The distribution function for all events of one topology is given by
$$
C(E_\nu)=\sum_i C_i(E_\nu) \, ,
$$
where $i$ runs over all events from a dataset,
and the number of events assigned to the $j$th bin is
$$
N_\nu^j=\int_{E_j}^{E_{j+1}} C(E_\nu) dE_\nu \, .
$$
 We assume that the uncertainty on $N_\nu^j$ is $\sqrt{N_\nu^j}$ when $N^j$ is large; when $N$ is small, namely smaller than one, we set a 90\% C.L. limit to the expected number of $N_\nu^j$ using  Poissonian statistics. 

Thanks to the effective areas (HESE \cite{Aartsen:2013jdh} and through-going muons \cite{Aartsen:2014cva}) it is possible to convert the $N_\nu^i$ into the expected flux in each energy interval, using the following formula for shower-like HESE
\begin{equation}
\frac{d\phi_{e,\tau}}{dE_\nu}(\bar{E}) = \frac{N_\nu^j}{\int_{E_j}^{E_{j+1}} (A_{\text{eff}}^e + (1-\eta) A_{\text{eff}}^\mu + A_{\text{eff}}^\tau) \ E^{-2} \ dE_\nu}   \ , 
\label{equ:aeff}
\end{equation}
and the following one for track-like HESE or through-going muons (they differ for the effective area $A_{\text{ eff}}^\mu$):
\begin{equation}
\frac{d\phi_{\mu}}{dE_\nu}(\bar{E}) = \frac{N_\nu^j}{\int_{E_j}^{E_{j+1}} \eta A_{\text{ eff}}^\mu \ E^{-2} \ dE_\nu}    \, ,
\end{equation}
where 
$$
\bar{E}=10^{(\log_{10}(E_j)+\log_{10}(E_{j+1}))/2}
$$ 
is the middle point on a log-scale, and $\eta=0.8$ represents the fraction of the $\nu_\mu$ effective area that is connected to track-like events (see~\cite{Palladino:2015zua}). The effective areas are taken from \cite{Aartsen:2013jdh} for HESE and from \cite{Aartsen:2014cva} for TGM.

The reconstructed data points are, for example, shown in \figu{recpoints}. Here the  uncertainties on the flux are proportional to the uncertainties on $N_\nu^j$ as described above, whereas the uncertainties on the energy axis cover the interval between $E_j$ and $E_{j+1}$. 

In the previous calculation we have assumed an $E^{-2}$ spectrum in each energy interval between $E_j$ and $E_{j+1}$. This approximation is not accurate when we analyze low energies (especially for track-like events), in which the atmospheric neutrino spectrum $E^{-3.7}$ dominates. A possible solution is to choose small energy intervals to reduce this effect. With our assumption the background related to $\phi_\mu$ might be underestimated in the data points at very low energies.

\section{Tables of the probabilites}
\label{app:prob}

\begin{figure*}[t]
\caption{Table of the HESE showers. For each event the deposited energy, the reconstructed energy (smaller than indicated value within 67\% C.L.), the direction and the probability that it comes from the different components are given. Probabilities larger than 30\% are marked in yellow.} 
\label{tab:shower}
\centering
\includegraphics[scale=0.45]{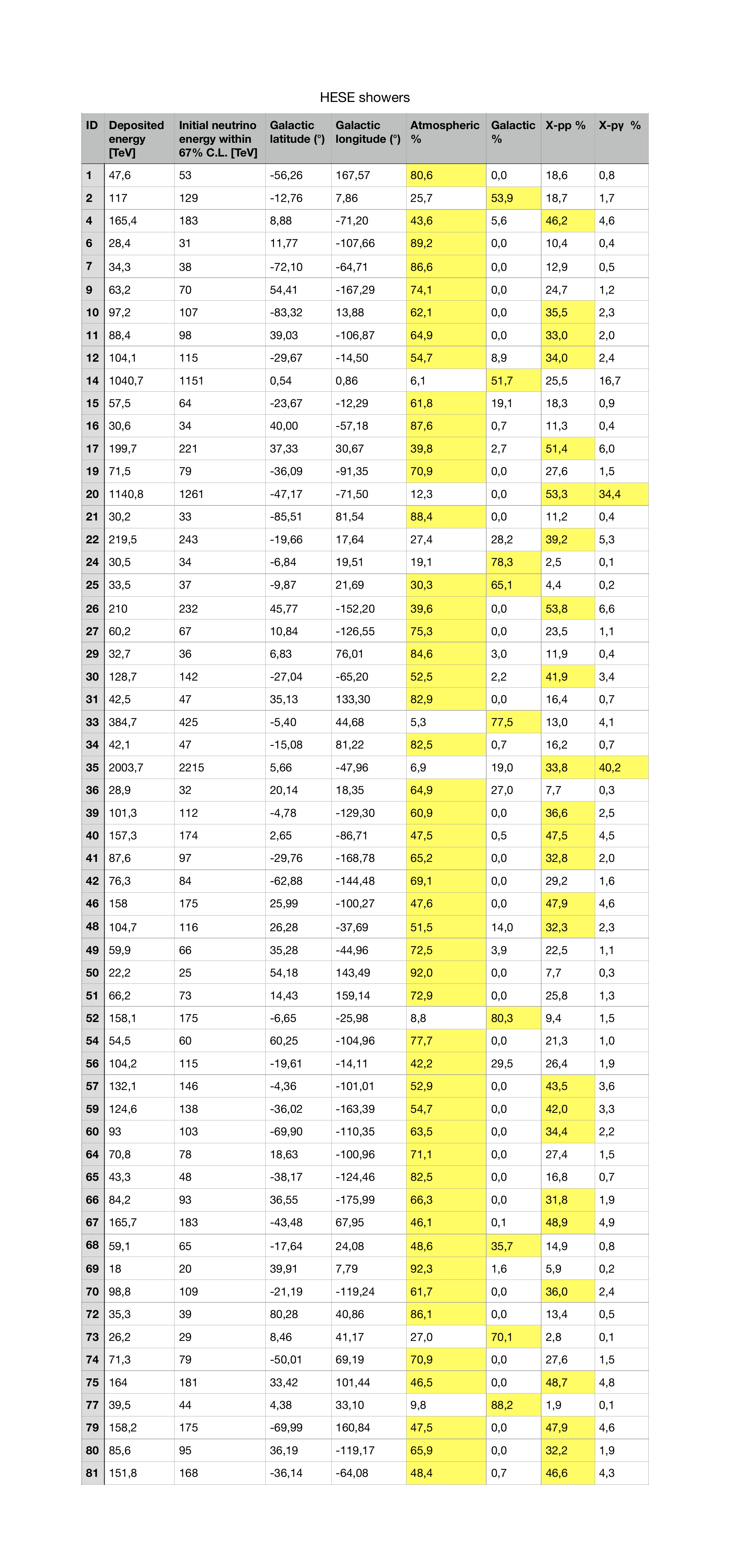}
\end{figure*}

\begin{figure*}[t]
\caption{Table of the HESE tracks. For each event the deposited energy, the reconstructed energy (smaller than indicated value within 67\% C.L.), the direction and the probability that it comes from the different components are given. Probabilities larger than 30\% are marked in yellow.}
\label{tab:track}
\centering
\includegraphics[scale=0.7]{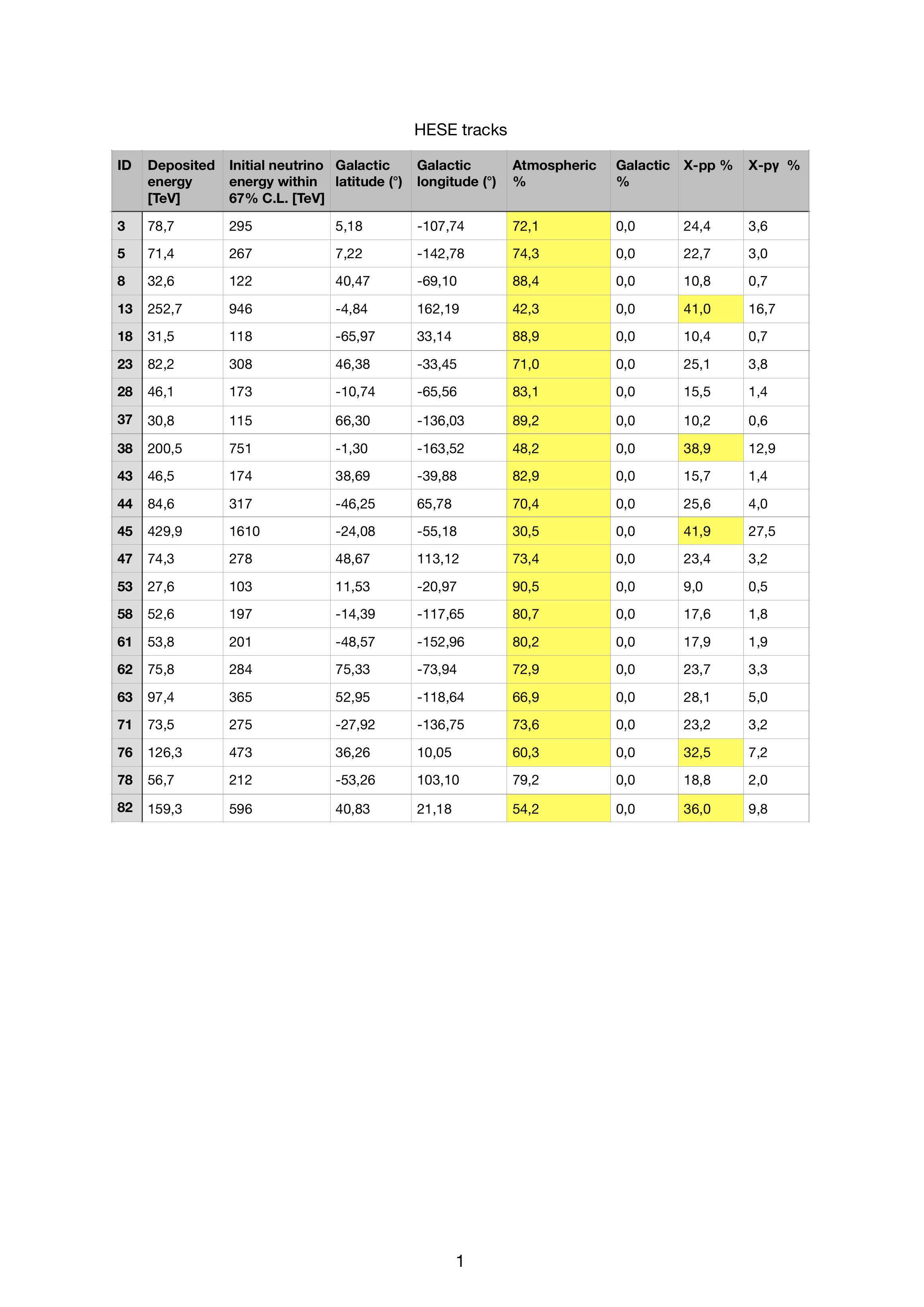}
\end{figure*}

\begin{figure*}[t]
\centering
\caption{Table of the through-going muons. For each event the deposited energy, the reconstructed energy (smaller than indicated value within 67\% C.L.), the direction and the probability that it comes from the different components are given. Probabilities larger than 30\% are marked in yellow.}
\label{tab:muon}
\includegraphics[scale=0.7]{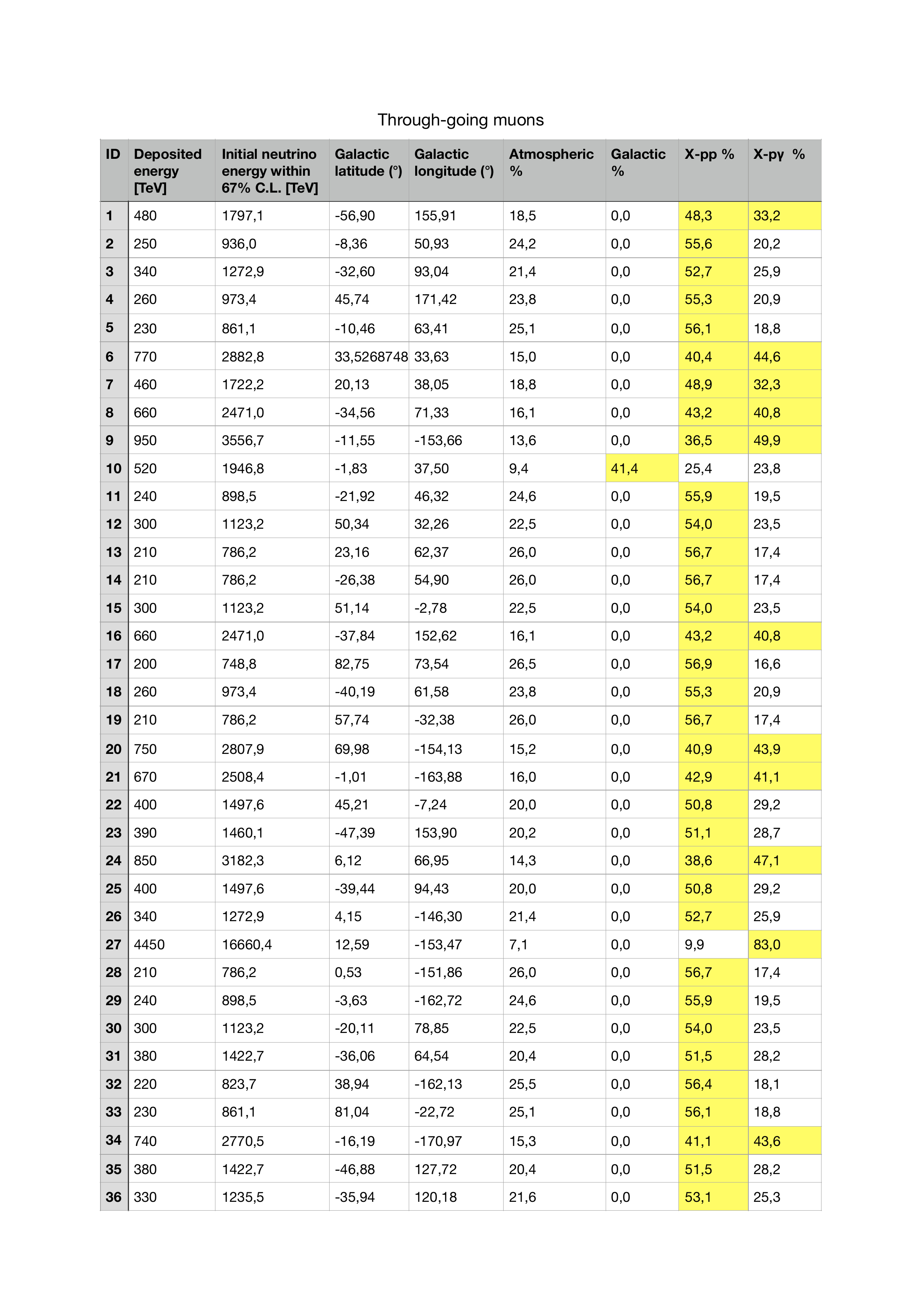}
\end{figure*}

In this Appendix list the event-to-event probability for each IceCube event for our multi-component model. The probabilities have been obtained using \equ{prob}; see main text.   We show three  different tables for HESE showers (\Tab~\ref{tab:shower}), HESE tracks (\Tab~\ref{tab:track}) and TGM (\Tab~\ref{tab:muon}).

\end{document}